\documentclass[aps,reprint,superscriptaddress,showpacs,showkeys,amsmath,amssymb,floatfix,nofootinbib]{revtex4-1}
\usepackage[colorlinks=true,citecolor=blue,filecolor=blue,linkcolor=blue,urlcolor=blue,pdftex]{hyperref}
\usepackage[usenames,dvipsnames]{color}
\usepackage{graphicx}
\usepackage{bm}
\usepackage[perpage]{footmisc}
\usepackage{amsmath,amsfonts,amssymb,wasysym,ifsym}
\usepackage{MnSymbol}
\usepackage{verbatim}
\usepackage[normalem]{ulem}
\usepackage{soul}
\usepackage{fancyhdr}
\usepackage{datetime}
\usepackage{multirow}
\usepackage{tabularx}
\usepackage{varwidth}
\usepackage{upgreek}

\newcommand{\inlinemaketitle}{{\let\newpage\relax\maketitle}}

\begin{document}

\newcommand{\bern}{\affiliation{Albert Einstein Center for Fundamental Physics, University of Bern, Bern 3012, Switzerland}}
\newcommand{\bologna}{\affiliation{Department of Physics and Astrophysics, University of Bologna and INFN-Bologna, Bologna 40126, Italy}}
\newcommand{\chicago}{\affiliation{Department of Physics \& Kavli Institute of Cosmological Physics, University of Chicago, Chicago, IL, USA}}
\newcommand{\coimbra}{\affiliation{Department of Physics, University of Coimbra, Coimbra 3004-516, Portugal}}
\newcommand{\columbia}{\affiliation{Physics Department, Columbia University, New York, NY 10027, USA}}
\newcommand{\lngs}{\affiliation{INFN-Laboratori Nazionali del Gran Sasso and Gran Sasso Science Institute, L'Aquila 67100, Italy}}
\newcommand{\mainz}{\affiliation{Institut f\"ur Physik \& Exzellenzcluster PRISMA, Johannes Gutenberg-Universit\"at Mainz, Mainz 55099, Germany}}
\newcommand{\heidelberg}{\affiliation{Max-Planck-Institut f\"ur Kernphysik, Heidelberg 69117, Germany}}
\newcommand{\munster}{\affiliation{Institut f\"ur Kernphysik, Wilhelms-Universit\"at M\"unster, M\"unster 48149, Germany}}
\newcommand{\nikhef}{\affiliation{Nikhef and the University of Amsterdam, Science Park, Amsterdam 1098XG, Netherlands}}
\newcommand{\nyuad}{\affiliation{New York University Abu Dhabi, Abu Dhabi, United Arab Emirates}}
\newcommand{\purdue}{\affiliation{Department of Physics and Astronomy, Purdue University, West Lafayette, IN 47907, USA}}
\newcommand{\rpi}{\affiliation{Department of Physics, Applied Physics and Astronomy, Rensselaer Polytechnic Institute, Troy, NY 12180, USA}}
\newcommand{\rice}{\affiliation{Department of Physics and Astronomy, Rice University, Houston, TX 77005, USA}}
\newcommand{\stockholm}{\affiliation{Oskar Klein Centre, Department of Physics, Stockholm University, AlbaNova, Stockholm SE-10691, Sweden}}
\newcommand{\subatech}{\affiliation{SUBATECH, Ecole des Mines de Nantes, CNRS/In2p3, Universit\'e de Nantes, Nantes 44307, France}}
\newcommand{\torino}{\affiliation{INFN-Torino and Osservatorio Astrofisico di Torino, Torino 10125, Italy}}
\newcommand{\ucla}{\affiliation{Physics \& Astronomy Department, University of California, Los Angeles, CA 90095, USA}}
\newcommand{\ucsd}{\affiliation{Department of Physics, University of California, San Diego, CA 92093, USA}}
\newcommand{\wis}{\affiliation{Department of Particle Physics and Astrophysics, Weizmann Institute of Science, Rehovot 7610001, Israel}}
\newcommand{\zurich}{\affiliation{Physik-Institut, University of Zurich, Zurich 8057, Switzerland}}


\author{E.~Aprile}\columbia
\author{J.~Aalbers}\nikhef
\author{F.~Agostini}\lngs\bologna
\author{M.~Alfonsi}\mainz
\author{F.~D.~Amaro}\coimbra
\author{M.~Anthony}\columbia
\author{F.~Arneodo}\nyuad
\author{P.~Barrow}\zurich
\author{L.~Baudis}\zurich
\author{B.~Bauermeister}\stockholm\mainz
% ...

\collaboration{XENON Collaboration}\email[E-mail: ]{xenon@lngs.infn.it}\noaffiliation

\title{XENON100 dark matter results from a combination of 477 live days}
%

\date{\today}
\begin{abstract}
We report on WIMP search results of the XENON100 experiment, 
combining three runs
summing up to 477 live days from January 2010 to January 2014. Data
from the first two runs were already published. A blind analysis was
applied to the last run recorded between
April 2013 and January 2014 prior to combining the results. The ultralow
electromagnetic background of the experiment,  
$\sim 5 \times 10^{-3}$~events/(keV$_{\mathrm{ee}}\times$kg$\times$day)
before electronic recoil rejection, together with the increased exposure 
of 48~kg~$\times$~yr improves the sensitivity. A profile
likelihood analysis using an energy range of $(6.6-43.3)$~keV$_{\mathrm{nr}}$ 
sets a limit on the elastic, spin-independent WIMP-nucleon scattering cross section for WIMP masses above 8~GeV/$c^2$, with a minimum of 1.1$\times 10^{-45}$~cm$^2$ at 50~GeV/$c^2$ and 90\% confidence level. We also report updated constraints on the elastic, spin-dependent
WIMP-nucleon cross sections obtained with the same data. We set upper
limits on the WIMP-neutron (proton) cross section with a minimum of 2.0$\times 10^{-40}$~cm$^2$ (52$\times 10^{-40}$~cm$^2$) at a WIMP mass of 50~GeV/$c^2$, at 90\% confidence level.
\end{abstract}



{\let\newpage\relax\maketitle}

\section{\label{sec:intro}Introduction}
Astrophysical observations at various scales
give strong evidence for
the existence of a nonluminous (rarely interacting), nonbaryonic,
and nonrelativistic (cold) matter component
that makes up 27\% of the total mass-energy budget
of the Universe, consisting of yet undetected particles whose nature
remains unknown \cite{Harvey:2015hha,Ade:2015xua}. Many theories beyond the Standard Model of particle
physics predict possible candidates, the most promising of which are
weakly interacting massive particles (WIMPs) 
\cite{Jungman:1995df,Bertone:1900zza}. In this paradigm, WIMPs
would interact with target nuclei of detectors placed deeply underground,
shielded by the rock overburden,
inducing detectable nuclear recoil (NR) signals.

A plethora of experiments worldwide are devoted to observing
the low-energy NRs of a few keV induced by WIMPs scattering off a nucleus~\cite{Undagoitia:2015gya}. Among these, the XENON100 experiment exploits a dual-phase (liquid-gas) xenon time projection chamber (TPC)~\cite{Aprile:2011dd}.
An electric ``drift'' field of $\sim$500 V/cm is applied across the
liquid xenon (LXe) volume by quasitransparent
electrodes (meshes); a stronger electric ``extraction'' field
of $\sim$12~kV/cm is applied in the gaseous xenon (GXe) multiplication region above the liquid-gas interface.

Particles interacting in LXe create a scintillation
light signal (S1) that is directly measured by 178 Hamamatsu R8520-AL photomultiplier tubes (PMTs), as
well as ionization electrons that can escape the local
ionization field and migrate along the drift field
direction towards the top of the TPC. Those ionization electrons
that reach the liquid-gas interface are extracted into
the GXe and accelerated by the extraction field
producing a scintillation signal (S2) that is proportional to the
number of extracted ionization electrons. The S1 and S2 signal timing and S2 hit pattern are used to determine the X,Y,Z coordinates of an interaction~\cite{Aprile:2011dd}. This event-by-event 3D-position information
can be used to define an optimal fiducial
volume to increase the signal to background ratio.

The XENON100 detector~\cite{Aprile:2011dd} features an active dark matter target of 62 kg and is installed at the Laboratori Nazionali del Gran Sasso (LNGS, Italy). Careful material selection~\cite{Aprile:2011ru} and detector design lead to very low backgrounds from electronic (ER)~\cite{Aprile:2011vb} and nuclear recoils (NR)~\cite{Aprile:2013tov}.
During the operation period between 2009 and 2016, three science runs (dark matter data sets) were collected. The results of the first two runs, referred to as run~I 
(100.9~live days in 2010)~\cite{Aprile:2011hi,Aprile:2011ts} and run~II 
(224.6~live days during 2011 and 2012)~\cite{Aprile:2012nq,Aprile:2013doa} were published 
and provided the best constraints on the spin-independent as well as on the spin-dependent WIMP-neutron cross section
at the time of publication. The final run 
(run~III) was taken between 2013 and 2014 (153.6 live days) and its results 
are published here for the first time in combination with the other two runs.

In this work, several improvements to the analysis and statistical interpretation 
are discussed in Sec.~\ref{sec:analysis}. The results 
of the spin-independent (SI) and spin-dependent (SD) combined analysis 
of all 477 live days of XENON100 dark matter science data 
are presented in Sec.~\ref{sec:results}.


\section{\label{sec:analysis}WIMP Search Data Analysis}

This paper includes the reanalysis of run~I and run~II data and the 
first analysis of run~III 
data, where each run corresponds to a data set with different 
detector settings and background levels. 
This section describes the general analysis procedure common to 
all three runs, emphasizing the modifications to the procedure 
reported in~\cite{Aprile:2012vw}. Section~\ref{sec:energy_scale} defines the energy 
scale for NRs. Section~\ref{sec:run_dep_quan} describes the 
operational differences between the three runs and run-dependent detector quantities. 
A detailed description of the 
data selection criteria and signal acceptance follows in 
Secs.~\ref{sec:cuts}~and~\ref{sec:acceptance}, respectively. The signal and 
background models are discussed in Secs.~\ref{sec:signal}~and~\ref{sec:background}. 
In Sec.~\ref{sec:likelihood}, the likelihood function used for the final 
statistical inference is described.

\subsection{\label{sec:energy_scale}Energy scale}

For a given energy deposition, the scintillation photons that reach the PMT 
photocathode may create photoelectrons (PEs) that are then amplified 
within the PMT. The probability of detecting such scintillation photons is, 
among other effects~\cite{Aprile:2011dd}, dependent on the interaction position due to changing solid angles 
with respect to the PMT arrays. Hence, a light collection efficiency (LCE) correction, dependent on the position, needs to be applied 
to the signal in order to achieve a uniform detector response at a given energy. 
The corrected signal (cS1) represents a spatially uniform response in the detector. Similarly, the measured S2 signal has 
a spatial dependence on the position both in the horizontal plane mainly due to 
warping of the top meshes~\cite{Aprile:2011dd} and in the vertical direction because of the finite electron 
lifetime caused by electronegative impurities in the LXe. Both effects can be quantified to achieve a position corrected 
signal, cS2. More details on signal corrections are provided 
elsewhere~\cite{Aprile:2011dd}.

The S1 and S2 signals provide information on the energy released by particles 
interacting in LXe. In this analysis, nuclear recoil processes are of greatest 
interest. For the direct scintillation signal, the relationship between the nuclear 
recoil energy $E_{\rm{nr}}$ and cS1 is given by (see~\cite{Plante:2011hw} and references 
therein):
\begin{equation}
\label{eq:enr_s1}
E_{\rm{nr}} = \frac{\rm{cS1}}{L_{\rm{y}}} \frac{1}{\mathcal{L}_{\mathrm{eff}}(E_{\rm{nr}})} \frac{S_{\rm{ee}}}{S_{\rm{nr}}},
\end{equation}
where 
$S_{\rm{ee}}= 0.58$ and $S_{\rm{nr}}= 0.95$ describe the scintillation quenching due to 
the electric field~\cite{Aprile:2006kx}, $L_{\rm{y}}$ is the detector-dependent light yield at 
122~keV$_{\rm{ee}}$ (electron recoil equivalent energy) shown in Table~\ref{tab:parameters}, and
$\mathcal{L}_{\mathrm{eff}}$ is the LXe relative scintillation efficiency. 
The parametrization and uncertainties of $\mathcal{L}_{\mathrm{eff}}$ as a
function of E$_{\rm{nr}}$ are based on existing direct measurements~\cite{Aprile:2011hi}.

For the S2 signal, the energy scale is given by (see~\cite{Aprile:2013teh} and 
references therein):
\begin{equation}
\label{eq:enr_s2}
E_{\rm{nr}} = \frac{\rm{cS2}}{Y}\frac{1}{Q_{\rm{y}}(E_{\rm{nr}})},
\end{equation}
where the secondary amplification factor $Y$ is determined from the detector response 
to single electrons~\cite{Aprile:2013blg}
and the parametrization of $Q_{\rm{y}}(E_{\rm{nr}})$ is taken from~\cite{Aprile:2013teh}. 
The corrected S2 observed by the bottom PMT array, cS2$_\mathrm{b}$, is used for 
the following analysis. 
In contrast to previous publications~\cite{Aprile:2011hi,Aprile:2012nq}, where the signal model was only modeled in S1, this analysis also incorporates the calculated S2 distribution based on the accurate simulation of the secondary scintillation signal of NRs~\cite{Aprile:2013teh}.

\begin{table*}[]
\centering
\caption{Detector and analysis parameters considered in each run \label{table1}}
\label{tab:parameters}
\begin{tabular}{|l|l|c|c|c|}
\hline
\multicolumn{2}{|l|}{}    & Run~I & Run~II & Run~III  \\ 
\hline
\multirow{2}{*}{\it Science Campaign} & Live days (d) & 100.9  & 223.1	   & 153.0   \\
                     & Period & 2010   & 2011-2012   & 2013-2014   \\ 
                     \hline
\multirow{5}{*}{\it Detector condition} 
	& Average electron lifetime ($\upmu$s) &294  $\pm$ 37  &519 $\pm$ 64  &720 $\pm$110  \\ 
	& $L_y$  (PE/keV) &  2.20 $\pm$ 0.09 & 2.28 $\pm$ 0.04 & 2.25 $\pm$ 0.03 \\
	& S2 amplification (PE/e$^-$) & 18.6 $\pm$ 6.6   &19.6 $\pm$ 6.9  &17.1 $\pm$ 6.4  \\
	& Extraction field in gas (kV/cm) & 11.89 $\pm$ 0.02 & 10.30 $\pm$ 0.01 & 11.50 $\pm$ 0.02 \\
	& Drift field (V/cm)  & 533 & 533 & 500 \\
    \hline
    \multirow{2}{*}{\it Calibration} 
    & $^{60}$Co, $^{232}$Th ER calibration in S1 range (events)  & 4116 & 15337 & 10469\\
    & $^{241}$AmBe NR calibration in S1 range (events) & 55423 & 25315 & 92226 \\
    \hline
    \multirow{7}{*}{\it Analysis} 
	& Low S1 threshold (PE) &  $3$ & $3$ & $3$ \\
	& High cS1 threshold (PE) &  $30$ & $30$ & $30$ \\
	& Low S2 threshold (PE) &  $300$ & $150$ & $150$ \\
	& Fiducial mass (kg) & 48 & 34 & 34\\
    & Total selected sample (events) & 929 & 402  & 346 \\ 
	& Expected background in benchmark ROI (events) & $3.9\pm0.5$ & $1.7\pm0.3$ & $1.0\pm0.2$ \\
	& Candidates in benchmark ROI (events) & 3 &  1 & 1 \\
        \hline

\end{tabular}
\end{table*}



\subsection{\label{sec:run_dep_quan}Detector operation}


Science data taken with different detector conditions must be corrected 
individually to avoid large systematic uncertainties. Therefore, the 
corrections for the measured quantities in each run are treated
separately  and the relevant differences are outlined below and
summarized in Table~\ref{table1}. 

For the analysis of the combined data, the light yield at
122~keV$_{\rm{ee}}$ does not change significantly among the
different data sets. The S2 signal corrections
are treated individually in each run. In particular the average
electron lifetime increases from an average of
$(294\pm 37)~\upmu$s in run~I to an average of $(720\pm110)~\upmu$s in run~III, 
while the exact time evolution during the runs is used in the correction. 
Small differences of a few
$\pm100$~V in the anode voltage and in liquid 
level result in different S2 amplification factors as shown in Table~\ref{table1}.
The gain values for
the PMTs are monitored on a weekly basis 
and an average value over the data taking period of each run
is used. The $^{nat}$Kr concentration is larger
in run~I (360 $\pm$ 70)~ppt~\cite{Aprile:2011hi} compared to run~II 
($19\pm4 )$~ppt~\cite{Aprile:2012nq}~and~III
($6\pm1 )$~ppt. It is measured, similarly to~\cite{Aprile:2012nq}, in extracted GXe samples  from the detector using ultrasensitive rare gas mass spectrometry~\cite{Lindemann:2013kna}.


The detector response to NR and ER is characterized by 
a $^{241}$AmBe ($\alpha,n$) source and $^{137}$Cs, $^{60}$Co, $^{232}$Th 
$\gamma$-sources, respectively~\cite{Aprile:2011dd}. 
The $^{241}$AmBe source and 
low energy Compton tail of the high-energy $\gamma$-sources, $^{60}$Co 
and $^{232}$Th, are used to determine the signal acceptances of the event 
selection. The latter is also used to model the background events 
caused by $\beta$ and $\gamma$-particles. The total number of events
for each calibration run after applying the selection described below is shown
in Table~\ref{table1}.

\subsection{\label{sec:cuts}Data selection}  

The event selection criteria for identifying single scatter events
are described in previously published 
results~\cite{Aprile:2011hi,Aprile:2012nq} as well as in a detailed 
publication on the analysis of the XENON100 data~\cite{Aprile:2012vw}. 
For this analysis, there is no change to the selection for run~I. However, in addition to the already presented event selection for run~II,
a few postunblinding cuts were developed to improve data quality and signal purity, 
described below. For run~III, due to similar detector conditions, 
the criteria from run~II were adopted and tuned while blinding the dark matter data in the relevant energy range. 

\begin{figure}
  \begin{center}
   \includegraphics[scale = 0.4]{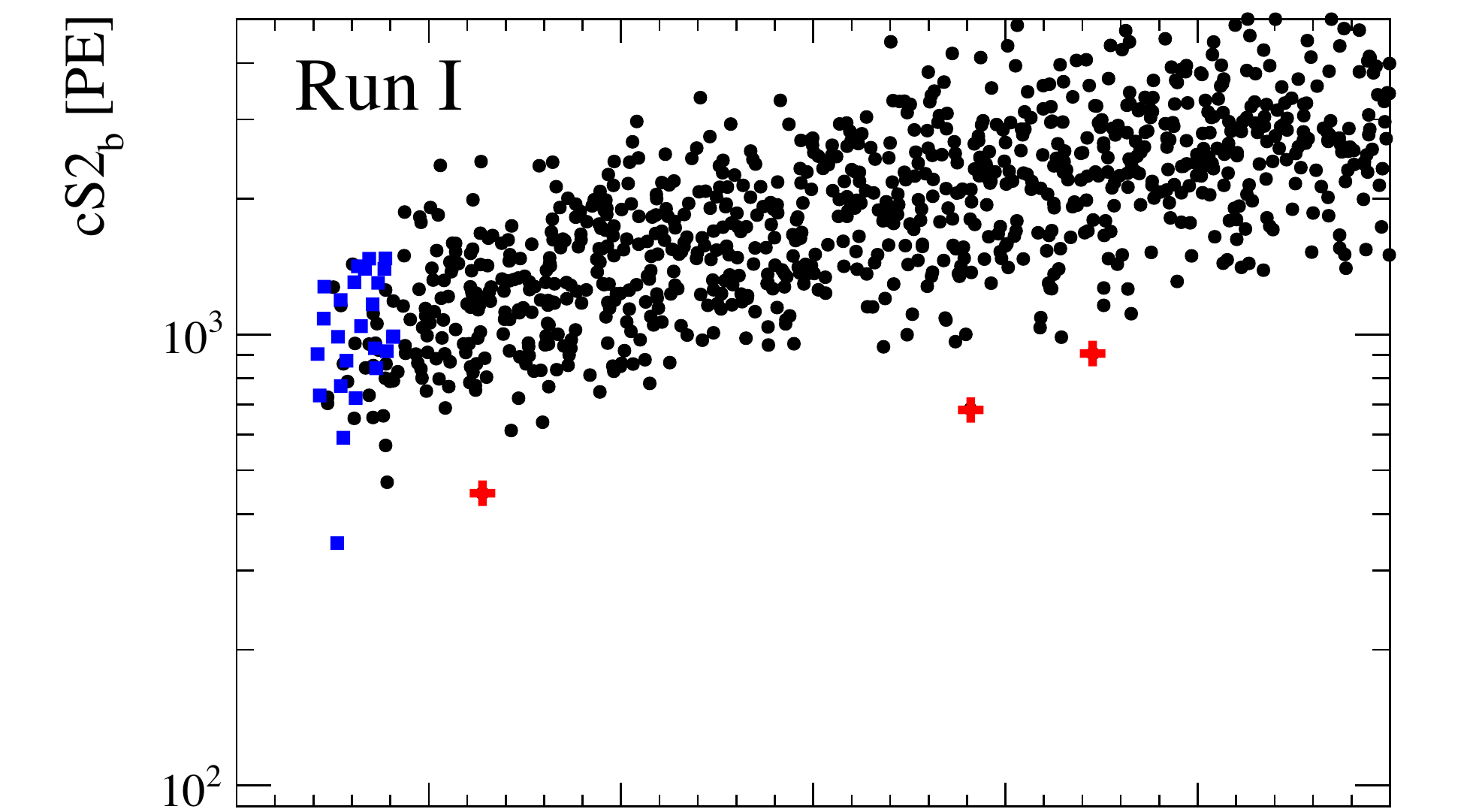}\vspace{-.02in}
   \includegraphics[scale = 0.4]{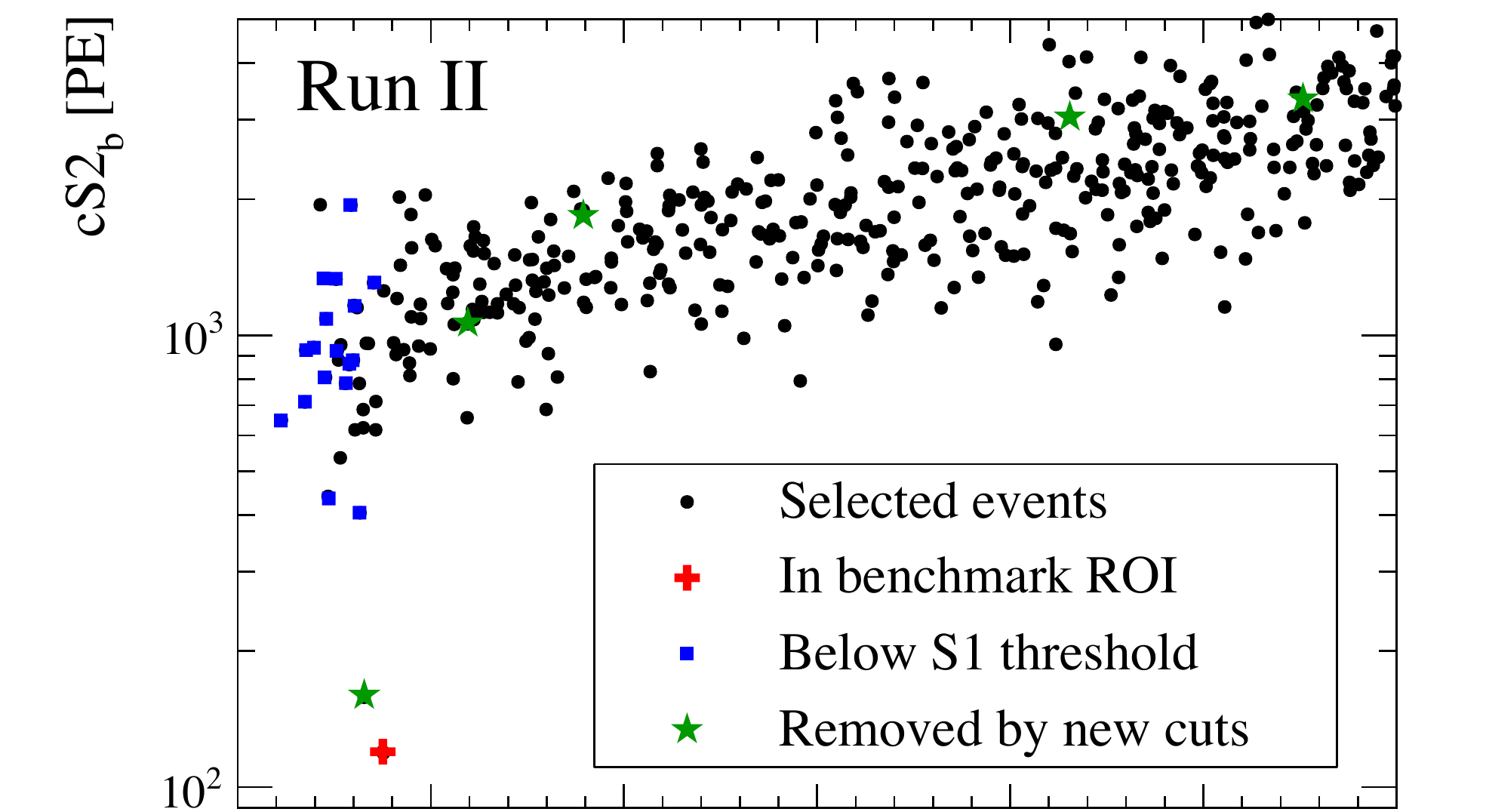}\vspace{-.02in}
   \includegraphics[scale = 0.4]{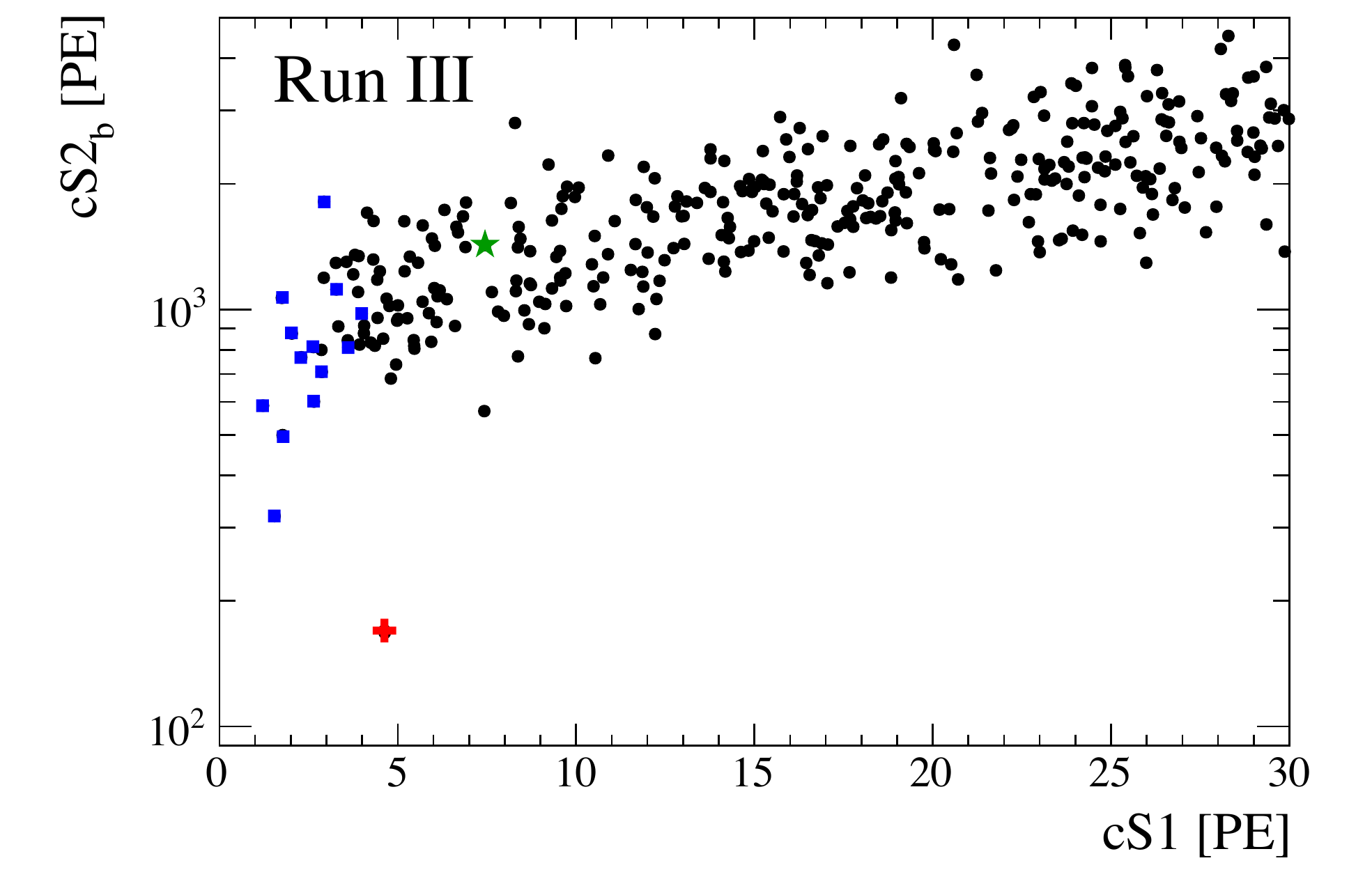}\vspace{-.1in}
   \caption[]{The cS1 and cS2$_{\rm b}$ for runs~I (top), II (middle), and III (bottom)
   science data passing all selection criteria (black circles, with red crosses for 
   dark matter candidates in the ROI). Events that fall below
   the S1 threshold (blue squares) are not used in the analysis. Events that 
   were removed by the new high S1 rate and improved S2 classification cuts 
   are also shown (green stars). The total number of events is summarized
   in Table~\ref{tab:parameters}.
     \label{Fig:data}}
  \end{center}	
\end{figure}

An analysis of the lone-S1 (an S1 without any correlated S2) rate over time 
revealed periods of significantly higher rates corresponding to a 
nonrandom occurrence of S1s. This increases the probability of 
an accidental coincidence with a lone random S2 in those periods, which could mimic 
the signature of a dark matter candidate event. The exact cause of this effect is not
known, but is indicative of unusual detector behavior and these time periods were 
excluded from the analysis. This new data quality criterion was 
optimized with the lone-S1 sample of the run~II dark matter data, removing
data periods where three or more lone S1s are present in a 500~second window. 
This data quality criterion was applied postunblinding to all runs. The 
optimization procedure, however, was fixed based on run II.
This criterion reduces the live times of runs~II~and~III by 
1.5~d and 0.6~d, respectively, and excludes one event from the run~II 
benchmark region as shown in Fig.~\ref{Fig:data}. No such high
rate periods were found in run~I.

A second novel selection criterion was determined from an improved 
S1 and S2 classification algorithm~\cite{j_aalbers_2016_58613}, 
initially developed for the next-generation 
experiment XENON1T~\cite{Aprile:2015uzo}. The new algorithm improves the identification 
of single electron S2s~\cite{Aprile:2013blg}, which the default XENON100 
algorithm sometimes misidentifies as an S1.
This new criterion has been applied postunblinding to run~II and
blinded to run~III dark matter data, reducing the expected non-Gaussian 
background (described in Sec.~\ref{sec:background}) by $\sim63\%$ with a signal
acceptance of $>98\%$ across the energy region of interest.

\begin{figure}[h]
  \begin{center}
   \includegraphics[scale = 0.4]{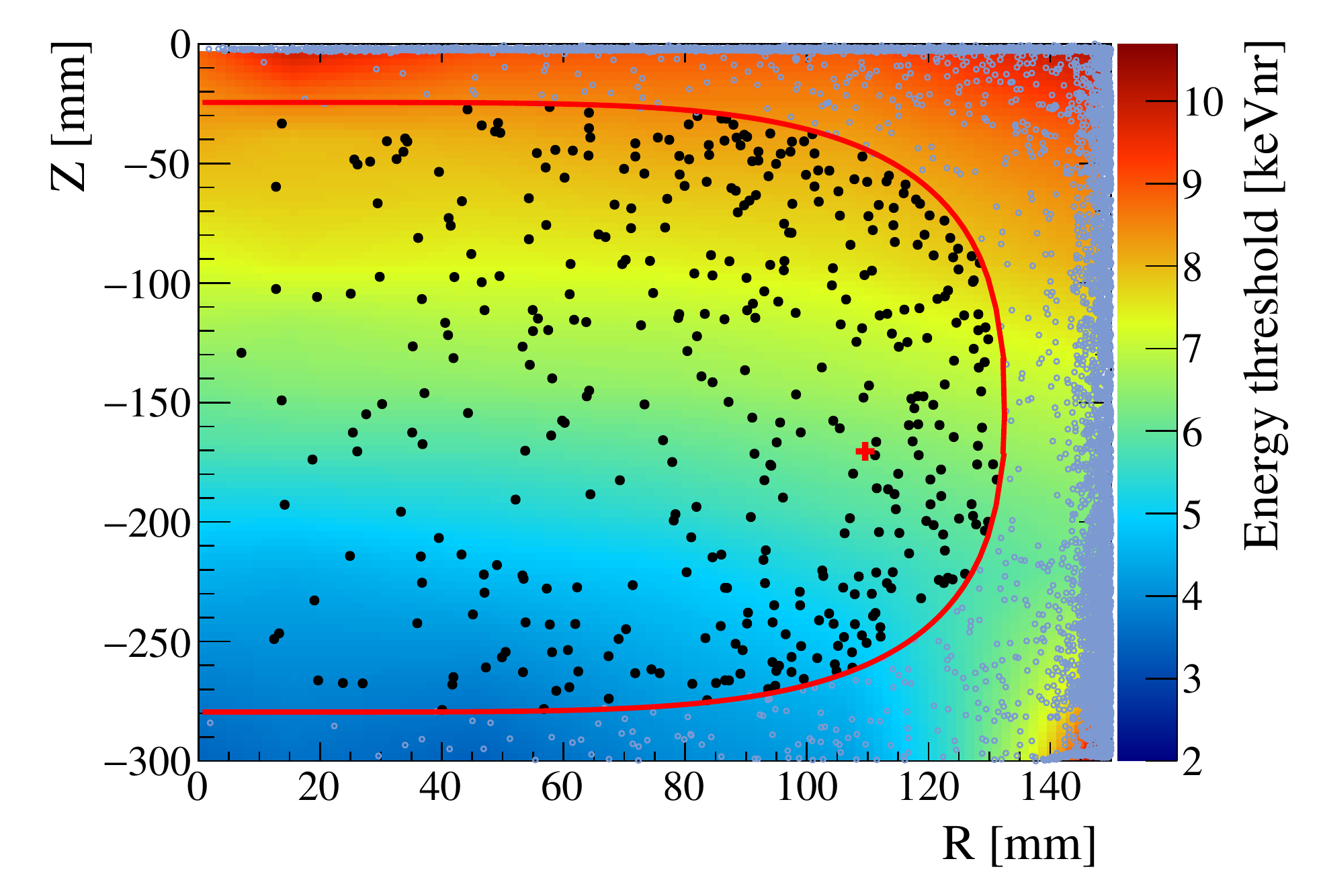}\vspace{-0.15in}
   \caption[]{The varying energy threshold
     in keV$_{\mathrm{nr}}$ due to the new threshold in S1,
     inside the active volume of the TPC as a function of the
     radius, R, and depth, Z. The color 
     scale is a mapping of the LCE and S1 $=$ 3~PE to energy assuming average 
     values of $\mathcal{L}_{\rm{eff}}$ and $Q_{\rm{y}}$. The run~III data
     inside (solid black points, with red cross for the candidate in the ROI) 
     and outside (hollow gray points) the fiducial volume (red line) are shown.
     \label{Fig:threshold}}
  \end{center}	
\end{figure}
\begin{figure}[h]
  \begin{center}
   \includegraphics[scale = 0.4]{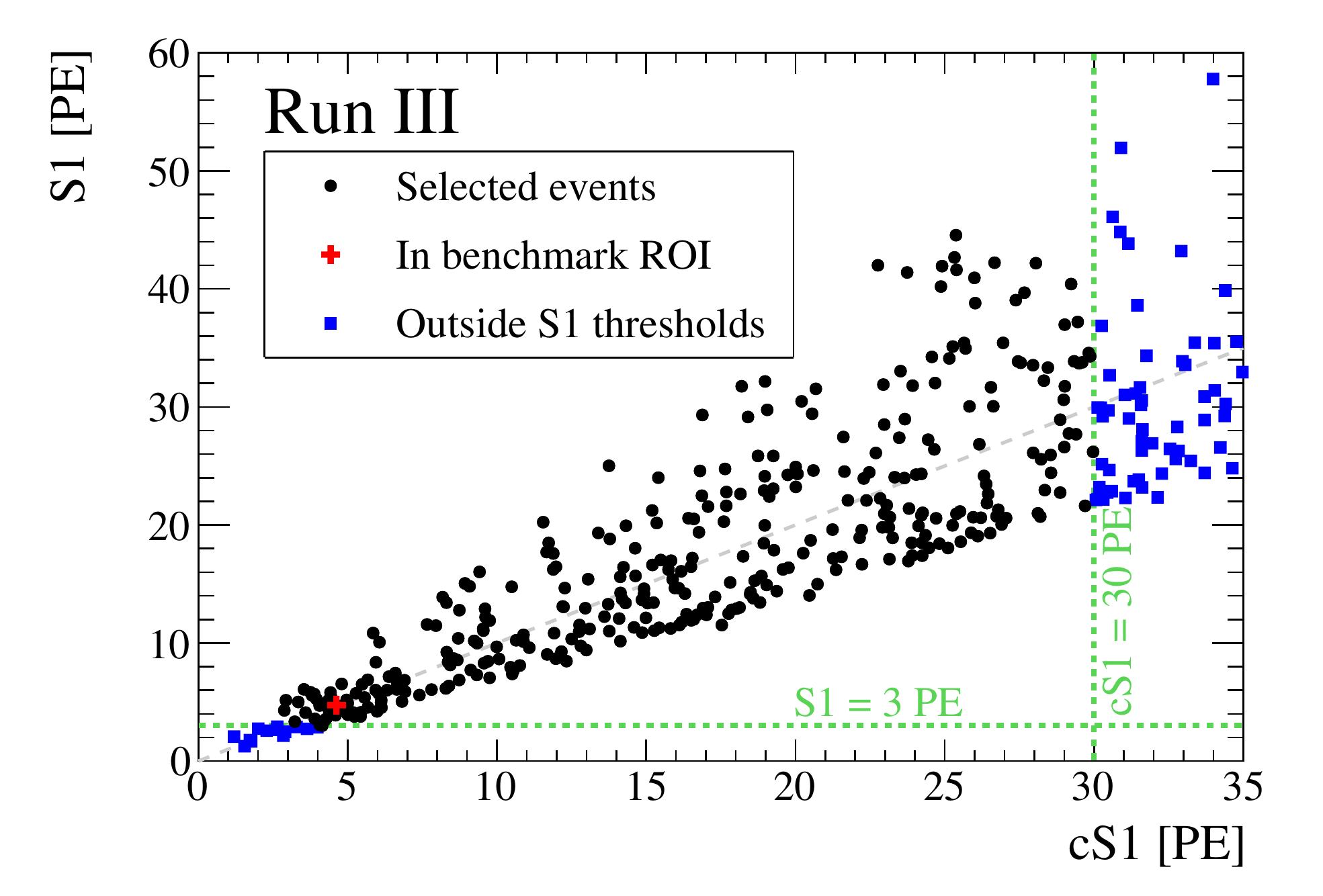}\vspace{-0.15in}
   \caption[]{The S1 and cS1 derived from the LCE for the run~III data. Events 
   below the lower S1 threshold at S1 = 3~PE (horizontal line) and 
   above the upper 
   threshold at 30~PE in cS1 (vertical line) are removed from the analysis (blue squares).
   The S1 = cS1 line is shown for reference.
   The dark matter candidate in the ROI is highlighted (red cross). 
   \label{Fig:us1_vs_cs1}}
  \end{center}	
\end{figure}

For a 100\% S2-trigger efficiency in run~I, the threshold on
the minimum amplitude of the proportional scintillation signal
was set to S2~$>$~300 PE since the trigger roll off begins at 280~PE 
(see Fig.~2 of \cite{Aprile:2012vw}). Due to a lower 
trigger threshold in runs~II and III, the S2 threshold condition 
was improved to S2~$> 150$~PE. The S1 threshold is now equalized 
for all runs to S1~$> 3$~PE, while an upper limit of the S1 range 
is set on the corrected signal to cS1~$< 30$~PE. Since the probability to 
detect a signal depends on the number of photons produced at the interaction site prior 
to LCE corrections, using S1 instead of cS1 for the low 
energy threshold is a more proper treatment, 
which is especially important towards very low energies. This results 
in a variable energy threshold as shown in 
Fig.~\ref{Fig:threshold} where regions of the 
TPC with a higher LCE close to the bottom PMT array exhibit a smaller 
energy threshold 3~keV$_{\rm{nr}}$ (nuclear recoil equivalent energy), 
while the top region of the 
fiducial volume requires a minimum energy deposition of 8.5~keV$_{\rm{nr}}$.
The relation between S1 and cS1, after applying the LCE correction, and
the corresponding thresholds are shown in Fig.~\ref{Fig:us1_vs_cs1}.

The final dark matter candidate samples after selection are shown in 
Fig.~\ref{Fig:data} for runs~I,~II,~and~III, where the events removed
by the new selection criteria are highlighted. The y-axes of the plots are shown in units of the corrected S2 signal in the bottom array only (cS2$_{\rm b}$).
A benchmark region of interest (ROI) can be defined 
similarly to~\cite{Aprile:2012vw} after all selection criteria, 
between the upper and lower thresholds in cS1 and S1 as stated in 
Table~\ref{table1}. This ROI is bounded in cS2$_{\rm{b}}$/cS1 discrimination space 
above by the 99.75\% ER rejection line and below by the lower 3$\sigma$ 
quantile of the AmBe neutron calibration data. 

\subsection{\label{sec:acceptance} Signal acceptance \label{sec:acc}}

The signal acceptance is estimated similarly to~\cite{Aprile:2012vw} 
by defining a control sample from calibration data using all the selection
criteria (cuts) except the one whose acceptance is to be estimated. 
NR calibration data are used for most of the 
cuts, while cuts that are more susceptible to noise (S1 coincidence and 
electronic noise cuts~\cite{Aprile:2012vw}) use ER calibration data, which span more of the science data taking runs. 

The acceptance for a given cut is evaluated as a function of the primary
parameter used in that cut, for example cS1 for the single S1 cut or 
cS2 for the S2 width cut. The acceptance of the S1 coincidence cut, previously a 
function of cS1 as in Fig.~3 of~\cite{Aprile:2012vw}, was found to vary by 
up to 15\% with changing LCE. Thus, we now parametrize this acceptance as a 
function of S1 instead.

\begin{figure}[h]
  \begin{center}
   \includegraphics[scale = 0.4]{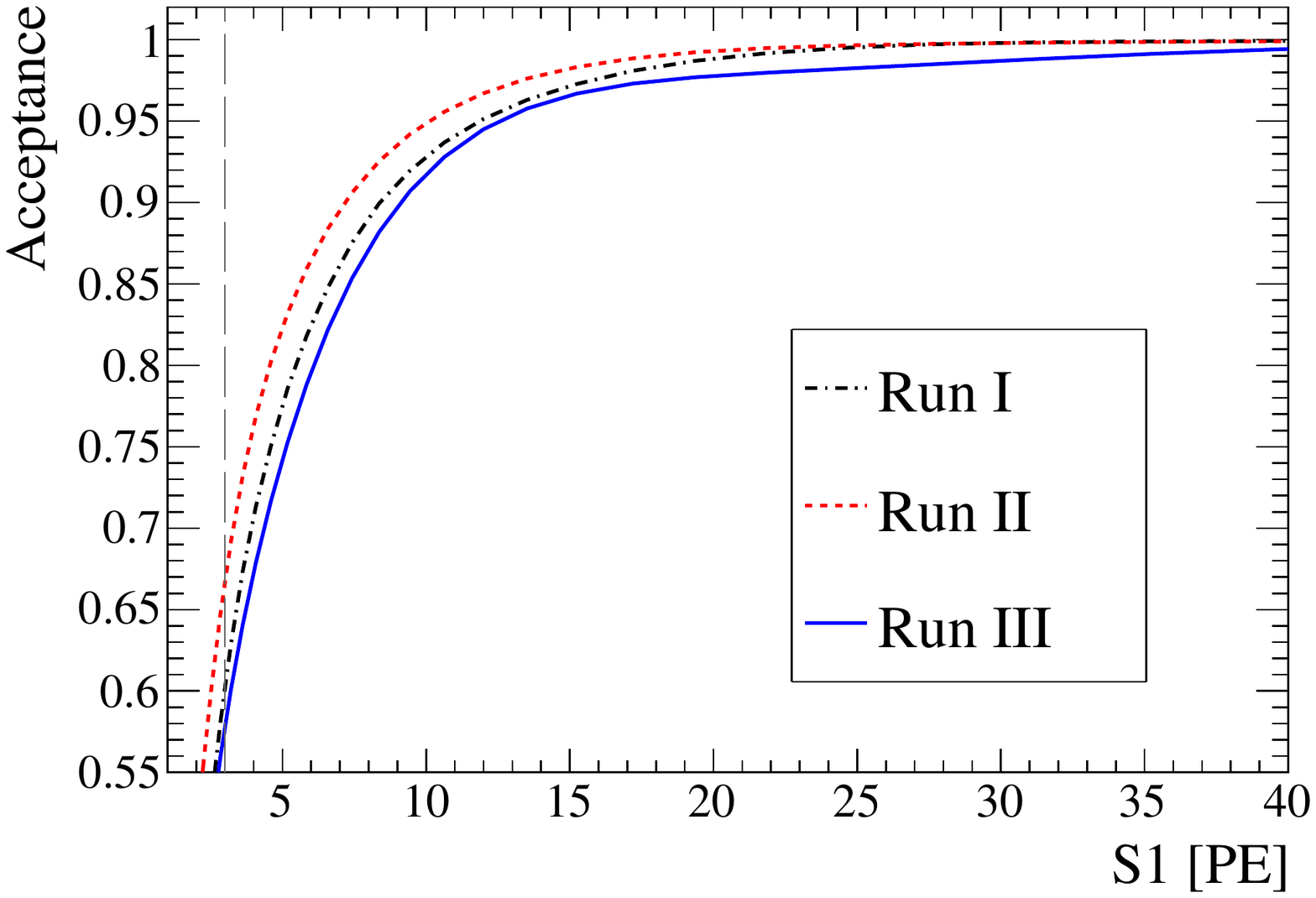}\vspace{-.1in}
   \includegraphics[scale = 0.4]{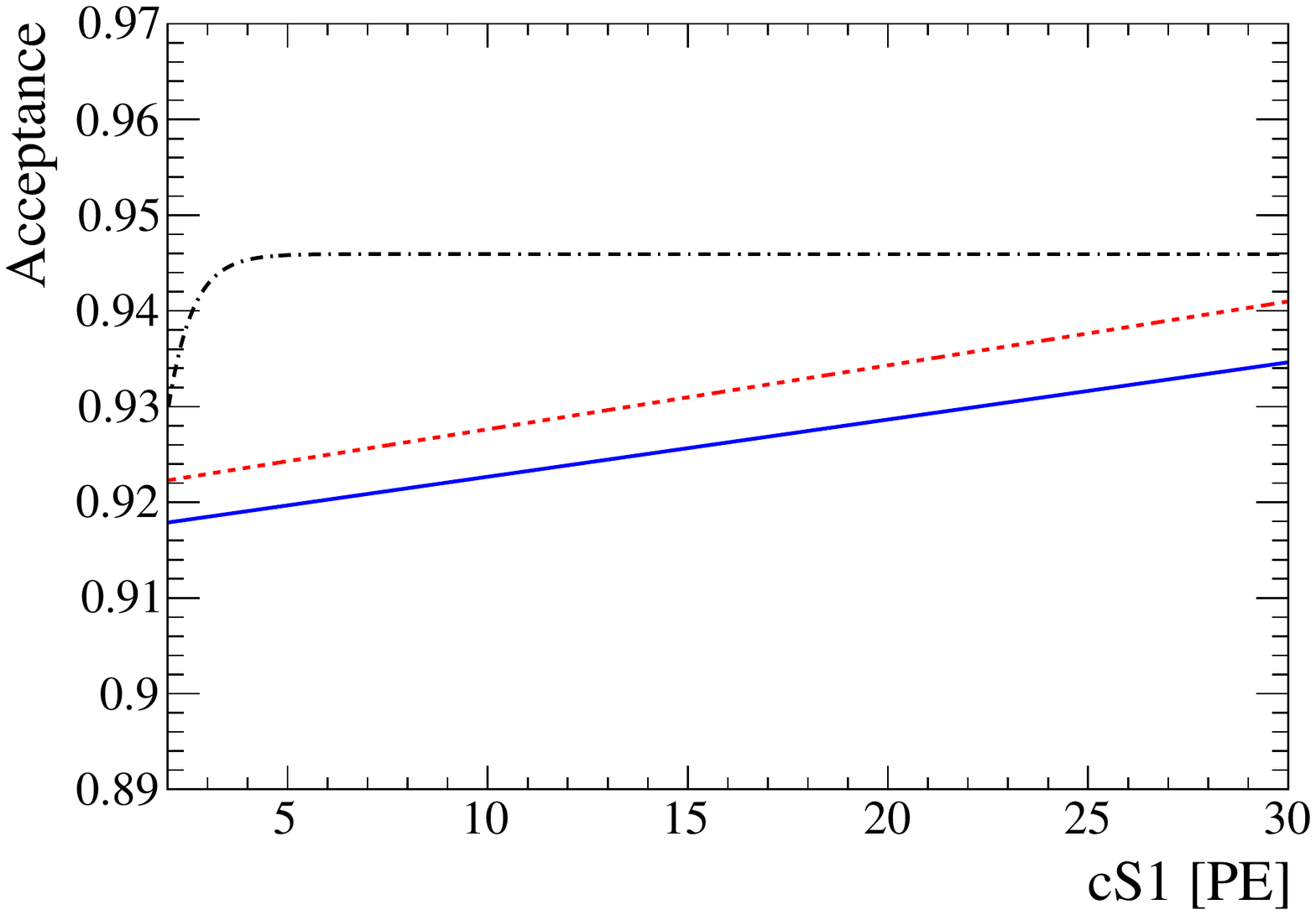}\vspace{-.1in}
   \includegraphics[scale = 0.4]{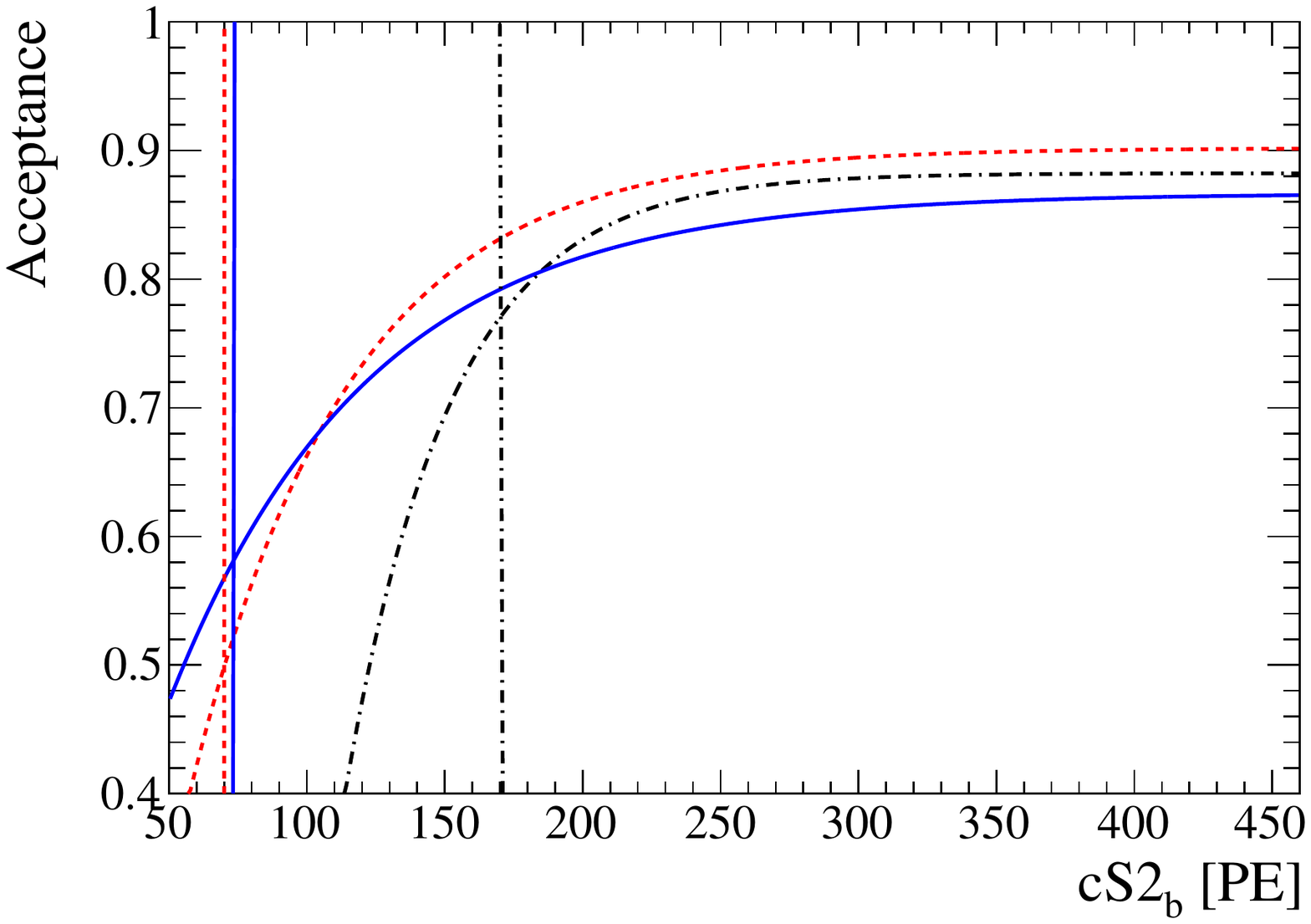}\vspace{-.1in}
   \caption[]{The combined acceptance of the S1 coincidence cut and noise misidentification (top), and the cumulative acceptance of the rest of the cS1-based cuts (middle) and cS2-based cuts (bottom), for each science run. 
   The acceptance in cS2 is constant above 450~pe. The S2 thresholds for each run are indicated by the vertical lines. \label{Fig:acceptance}}
  \end{center}	
\end{figure}

This analysis selects the primary S1 as that with the 
most PMT coincidences in a waveform. 
However, correlated electronic noise can be misidentified as the primary 
S1 and contaminate an event with a real signal, 
causing the event to be removed from the control samples 
and underestimating the acceptances. 
The acceptance loss is now estimated from the probability that a noisy 
peak accompanying a good S1 peak in an event is misidentified as the primary S1.
Figure~\ref{Fig:acceptance} (top) shows this new noise misidentification 
acceptance loss, combined with the S1 coincidence cut acceptance, 
as a function of S1.

The same procedure is applied across all three science runs and 
the cumulative acceptance of all the cuts in each parameter space is shown in 
Fig.~\ref{Fig:acceptance}. The small differences between runs 
are due to the varying detector parameters and cut optimization. 
The total uncertainty is estimated to be 
less than 20\% based on differences in $^{241}$AmBe or ER calibration data and 
the selection of the control samples. This increases the profile likelihood limit by 
a negligible few percent and is hence not considered as a nuisance parameter.
The acceptances of the S1, cS1 and S2 thresholds are taken 
into account by applying these cuts directly on the signal model, which is described in 
the following section. 

\subsection{\label{sec:signal} Signal model}

The signal model describing the rate of WIMP interactions, $R$, 
in the detector is given by~\cite{Lewin:1995rx}:
\begin{equation}
\frac{dR(m_{\chi}, \sigma)}{dE} = \frac{\rho_0}{m_{\chi}\cdot m_{\rm{A}}} \cdot \int v \cdot f(v) \cdot \frac{d\sigma}{dE} \left(E,v\right) d^3 v,
\label{eq:rate}
\end{equation}
where $E$ is $E_{\rm{nr}}$ in Eqs.~(\ref{eq:enr_s1})~and~(\ref{eq:enr_s2}), $\rho_0 = 0.3~\rm{GeV/cm^3}$ is the local dark matter density~\cite{Green:2011bv}, 
$m_{\chi}$ and $m_A$ are the WIMP and nucleus mass, respectively, 
and $f(v)$ is the distribution of dark matter particle velocities $v$. 
An isothermal WIMP halo is assumed for $f(v)$ with an escape velocity of 
$v_{esc} = 544~\rm{km/s}$~\cite{Smith:2006ym} and a local circular velocity of $v_0 = 220~\rm{km/s}$. 
The differential cross section, $\frac{d\sigma}{dE}$, is composed of 
a SI and SD contribution~\cite{Lewin:1995rx}:
\begin{equation}
\frac{d\sigma}{dE} = \frac{m_{\rm{A}}}{2\mu ^2 _A v^2} \cdot \left(\sigma_{SI} F^2 _{SI}(E) + \sigma_{SD} F^2 _{SD}(E) \right).
\end{equation}
where $\mu_A$ is the reduced mass of the nucleus and WIMP, and 
$F$ and $\sigma$ are the Helm form factors~\cite{Lewin:1995rx} 
and cross sections as $q \rightarrow 0$, respectively, for SI and SD interactions
described in the following sections. Each component is considered 
separately in the profile likelihood (PL) analysis below, 
with the other one being fixed to zero.

The rate as a function
of detector observables can then be written following~\cite{Aprile:2011hi} as
\begin{align}
\frac{d^2R(m_{\chi}, \sigma; \mathcal{L}_{\mathrm{eff}}, \mathrm{LCE}, Q_y)}{d(\mathrm{cS1})d(\mathrm{cS2_b})}  \approx  \epsilon(\mathrm{S1})\epsilon(\mathrm{cS1})\epsilon(\mathrm{cS2}_\mathrm{b}) \times \nonumber \\
\int\frac{dR}{dE} p(\mathrm{cS1}|E, \mathcal{L}_{\mathrm{eff}}, \mathrm{LCE}) p(\mathrm{cS2_b}|E, Q_y) dE,
\label{Eq:rate_obs}
\end{align}
where $\epsilon(\mathrm{S1})$, 
$\epsilon(\mathrm{cS1})$, and $\epsilon(\mathrm{cS2}_\mathrm{b})$ are the 
acceptances shown in Fig.~\ref{Fig:acceptance}, 
and $p$ denotes the probability
distribution functions (PDFs) to observe cS1 or cS2$_\mathrm{b}$ given 
a recoil energy, $E$~\cite{Aprile:2012nq}.
The approximation assumes a negligible anticorrelation between S1 
and S2 signals for NRs, as suggested by~\cite{Aprile:2013teh},
such that the acceptances
and probabilities can be multiplied independently as 
$p(\mathrm{cS1},\mathrm{cS2}) \approx p(\mathrm{cS1}) \cdot p(\mathrm{cS2})$.
Finally, the lower thresholds in S1 and S2 
are applied directly to the computed spectra, which 
can then be integrated to 
estimate the total number of expected signal events: 
\begin{equation}
\begin{split}
N_s&(m_{\chi},\sigma, \mathcal{L}_{\rm{eff}}, LCE, Q_{\rm{y}})= \\ 
& \int^{30}_{\mathrm{cS1}=0} {\int ^{\mathrm{cS2_b ^{up}}}_{\mathrm{cS2_b}=0}} \frac{d^2R
}{d(\mathrm{cS1})d(\mathrm{cS2_b})}d(\mathrm{cS1})d(\mathrm{cS2_b}),
\label{Eq:Ns}
\end{split}
\end{equation}
where $\mathrm{cS2_b^{up}}$ is an upper bound that includes the 
whole ER band.
The signal shape is given by the following PDF:
\begin{equation}
f_{s}(\mathrm{cS1},\mathrm{cS2_b}; m_{\chi}, \mathcal{L}_{\mathrm{eff}}, \mathrm{LCE}, Q_y) = 
\frac{1}{N_s} 
\frac{d^2R}{d(\mathrm{cS1})d(\mathrm{cS2_b})}.
\end{equation}
To account for uncertainties 
in the PL analysis below, the spectra are computed for each run, WIMP mass, LCE and values of $\mathcal{L} _{\rm{eff}}$ and $Q_{\rm{y}}$. 

\begin{figure}[h]
  \begin{center}
\hspace*{-0.5cm}                                                           
 \includegraphics[scale = 0.45]{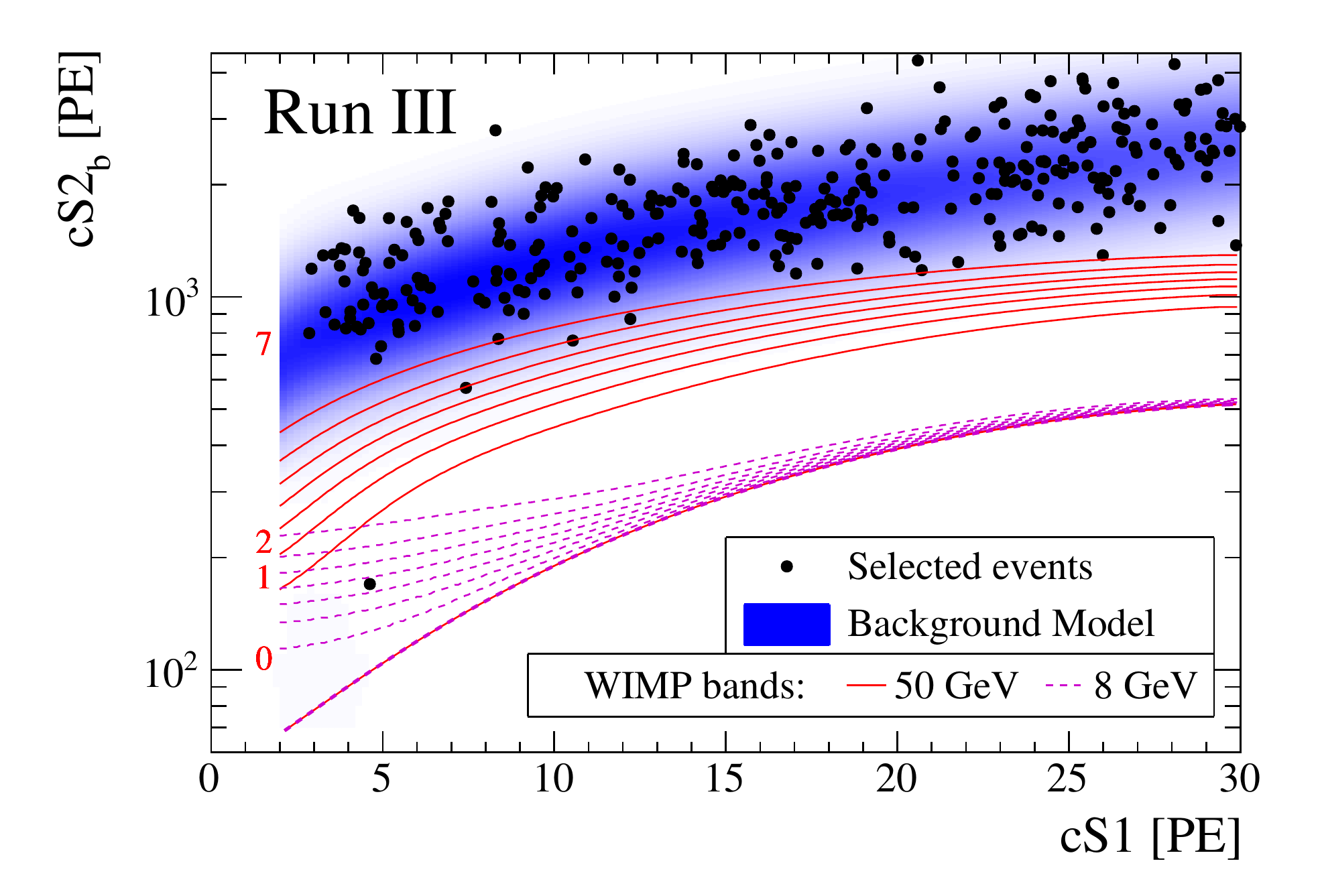}
    \caption[]{
    Example bands for
   8~GeV/$c^2$ (purple dashed lines) and 50~GeV/$c^2$ (red solid lines with numbered labels) 
   WIMP mass SI signal models. The lower bound for all WIMP masses is defined 
   by the 20~GeV/$c^2$ model as described in the text.
   The upper bound of the topmost band is beyond the vertical range.
   The shape of the background model 
   is shown with a (blue) linear color scale.
   The run~III science data are overlaid for reference.
   \label{Fig:pl_banding}}
  \end{center}	
\end{figure}

Following a similar procedure as in~\cite{Aprile:2011hx}
the (cS1, cS2$_\mathrm{b}$) spectra are binned
into 8 bands, with equal numbers of signal events in the nominal model,
to exploit the knowledge of the 
signal shape and allow the statistical interpretation in regions 
with optimal signal to background ratios. The lower bound is defined
by the 99.7\% acceptance line of the 20~GeV/$c^2$ WIMP signal model 
to keep the selected signal events for all WIMP masses fixed. The upper bound is defined
by $\mathrm{cS2_b^{up}}$ in Eq.~(\ref{Eq:Ns}).
Two examples of the banding 
are shown in Fig.~\ref{Fig:pl_banding}.

\subsubsection{Spin-independent cross section}
\begin{figure}[h]
  \begin{center}
   \includegraphics[scale = 0.4]{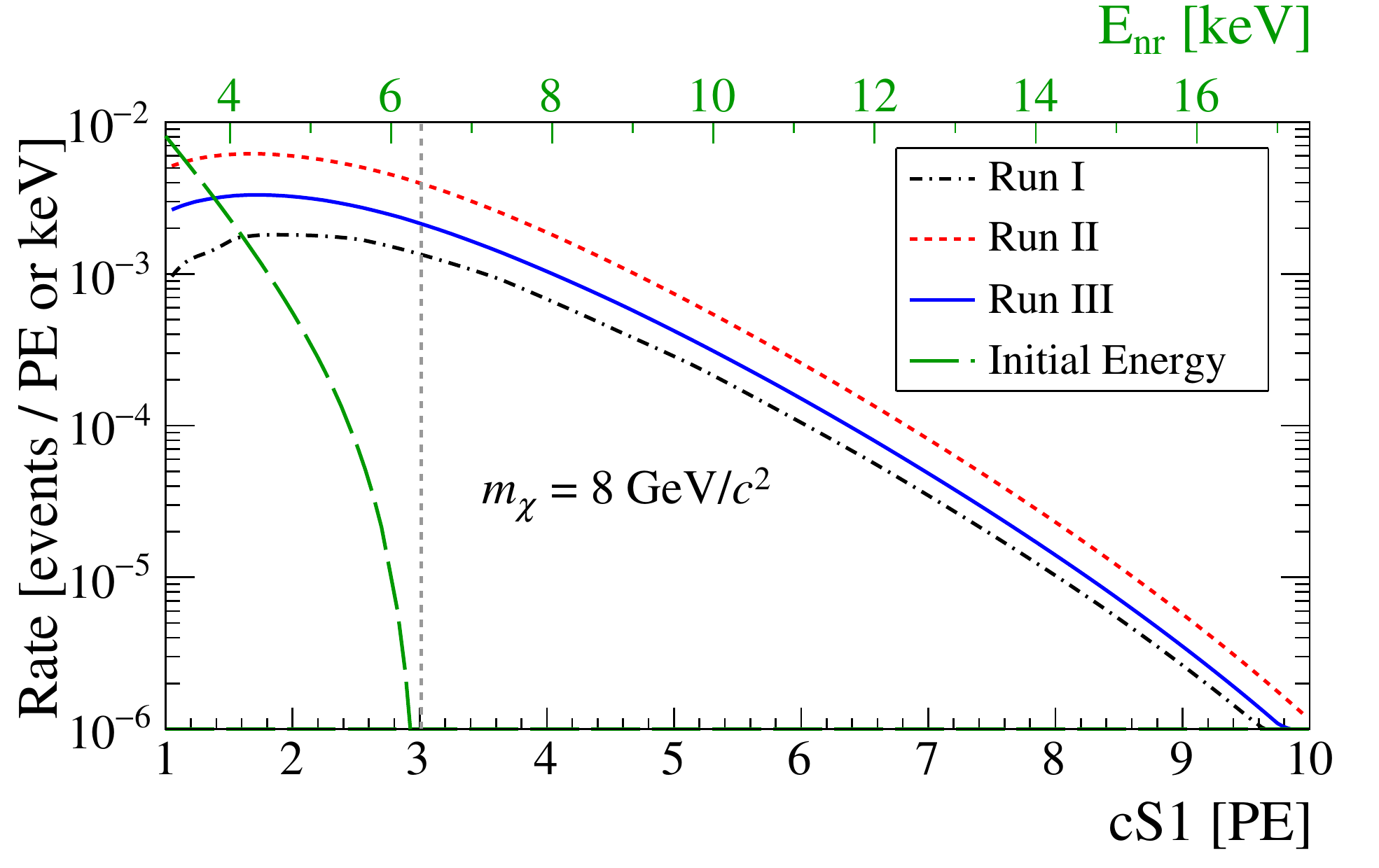}
   \includegraphics[scale = 0.4]{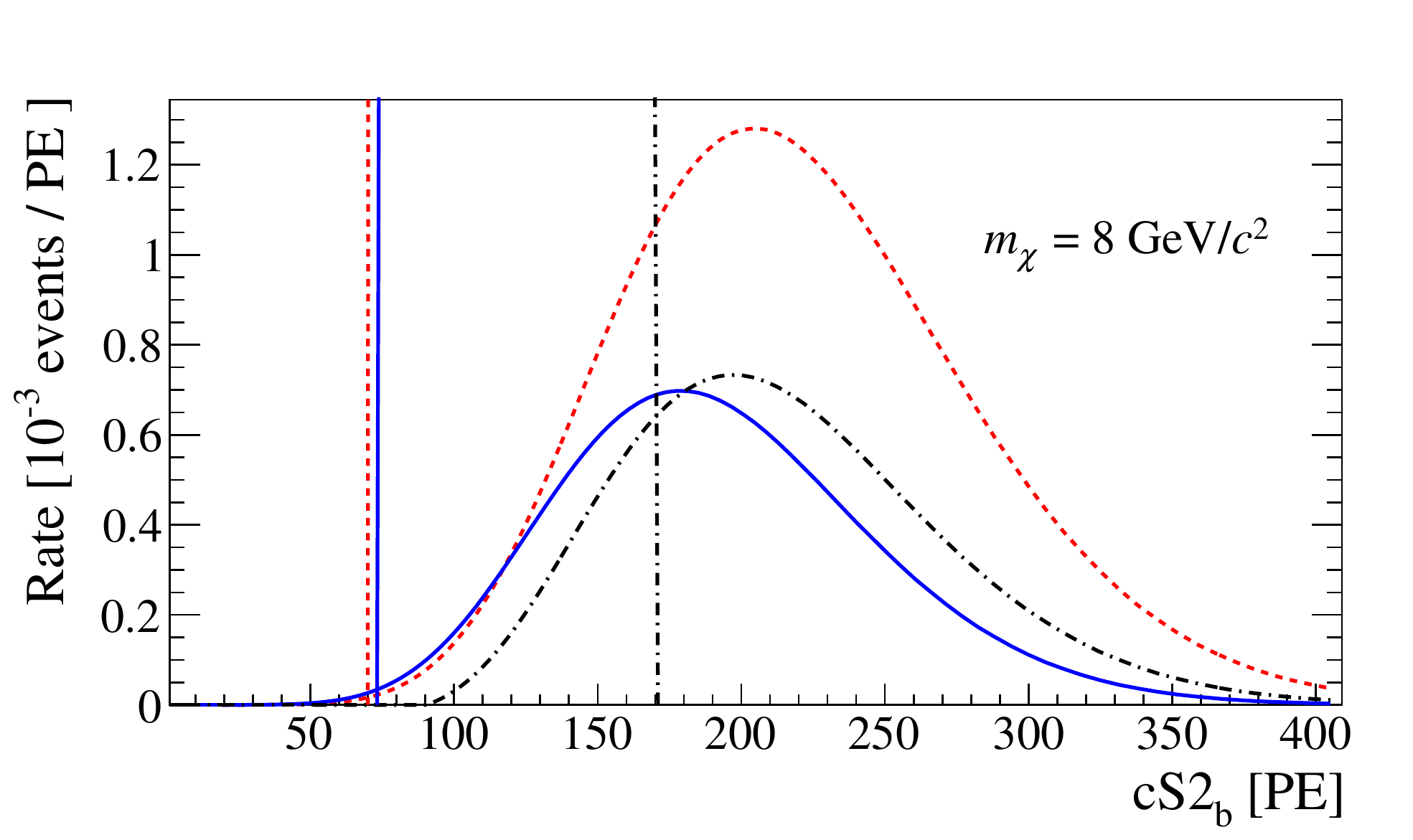}
   \caption[]{Expected rate in cS1 (top) and cS2$_{\rm b}$ (bottom) of an 8~GeV/$c^2$ WIMP SI 
   signal and a cross section of 1$\times 10^{-44}$~cm$^2$ for each science run. The initial
     energy spectrum in keV$_{\rm{nr}}$ is shown (green long-dashed).
     The average LCE = 1 is shown for example, 
   resulting in the hard cut at S1 = cS1 = 3 PE. The S2 thresholds for each run
   are indicated by the vertical lines. The differences in the spectra are due
   to varying exposures, S2 threshold, and acceptances. 
     \label{Fig:signal}}
  \end{center}	
\end{figure}

By assuming a spin-independent and isospin conserving interaction, 
the cross section can be computed as:
\begin{equation}
\sigma_{SI} = \sigma _{\rm p} \cdot \frac{\mu ^2 _{\rm A}}{\mu ^2 _{\rm p}} \cdot A^2,
\label{eq:SI}
\end{equation}
where $\sigma _{\rm p}$ is the WIMP-proton cross section, $A$ is the nucleus mass number and $\mu _p$ is the reduced 
mass of the proton and WIMP. Examples of corresponding computed spectra for each run
are shown in Fig.~\ref{Fig:signal}.
The green line in the top panel of Fig.~\ref{Fig:signal} is the 
energy spectrum as given by Eqs.~(\ref{eq:rate})~and~(\ref{eq:SI}) for an 8~GeV/$c^2$ WIMP. 
The observable cS1 and cS2$_\mathrm{b}$ spectra from Eq.~(\ref{Eq:rate_obs}) are also shown 
for each run, illustrating that for low WIMP masses, Poisson 
fluctuations of the generated signal 
quanta are essential to observe signals above the energy threshold of the detector.

\subsubsection{Spin-dependent cross section}
Following the work of~\cite{Aprile:2013doa}, a combination of the three science runs 
can also strengthen the dark matter spin-dependent interaction results. The corresponding 
structure functions are based on a chiral effective field theory  
considering two body currents as computed in~\cite{Menendez:2012tm}, resulting in 
the following cross section:
\begin{equation}
\sigma_{SD} = \frac{32}{\pi} \mu ^2 _A \cdot G_F ^2 [a_p \langle S_p \rangle + a_n \langle S_n \rangle]^2 \cdot \frac{J+1}{J},
\end{equation}
where $G_F$ is the Fermi coupling constant, $J$ is the the total nuclear spin, 
$a_{p,n}$ are the effective proton and neutron couplings, and $\langle S_{p,n} \rangle$
is the expectation of the total nuclear spin operator.

\subsection{\label{sec:background} Background model}

This section describes how the ER and NR backgrounds 
are modeled and combined into a total background model. 
These are derived similarly to 
the run~II method in~\cite{Aprile:2012nq} with the addition of a 
new method to model the accidental coincidence component of the 
ER non-Gaussian background. 

\begin{figure}[h]
  \begin{center}
    \includegraphics[scale = 0.45]{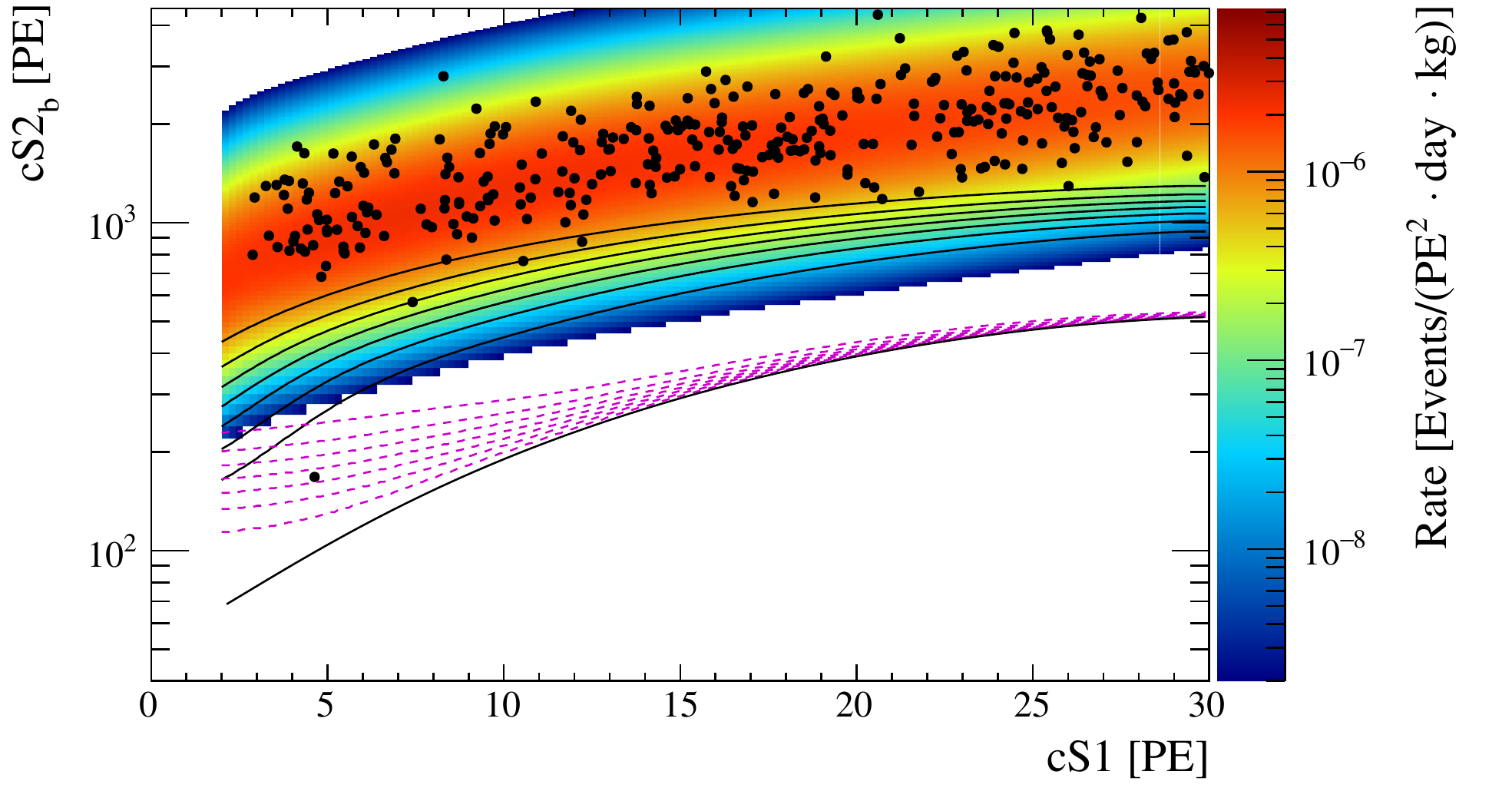}\vspace{-.31in}
    \includegraphics[scale = 0.45]{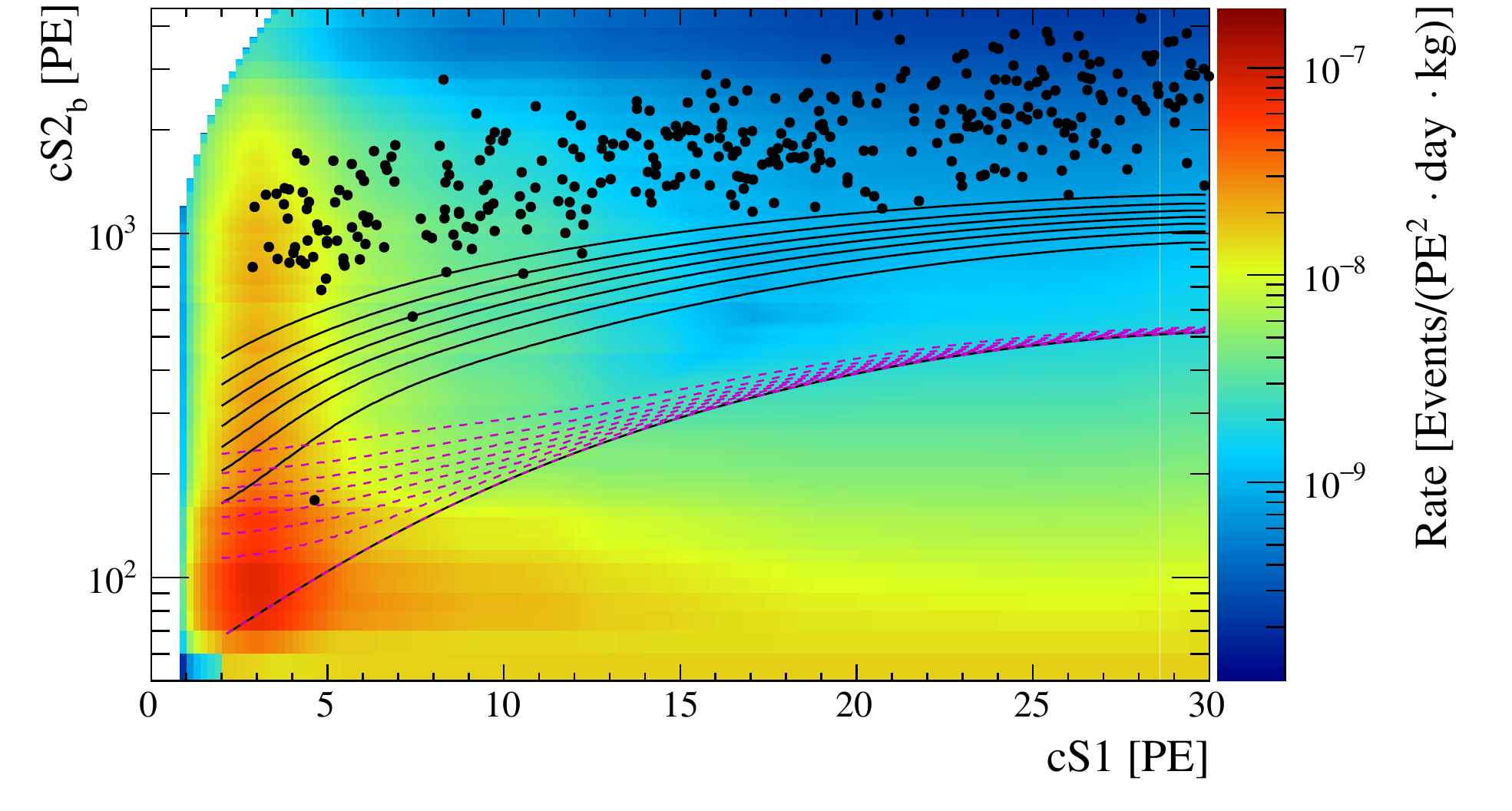}\vspace{-.31in}
    \includegraphics[scale = 0.45]{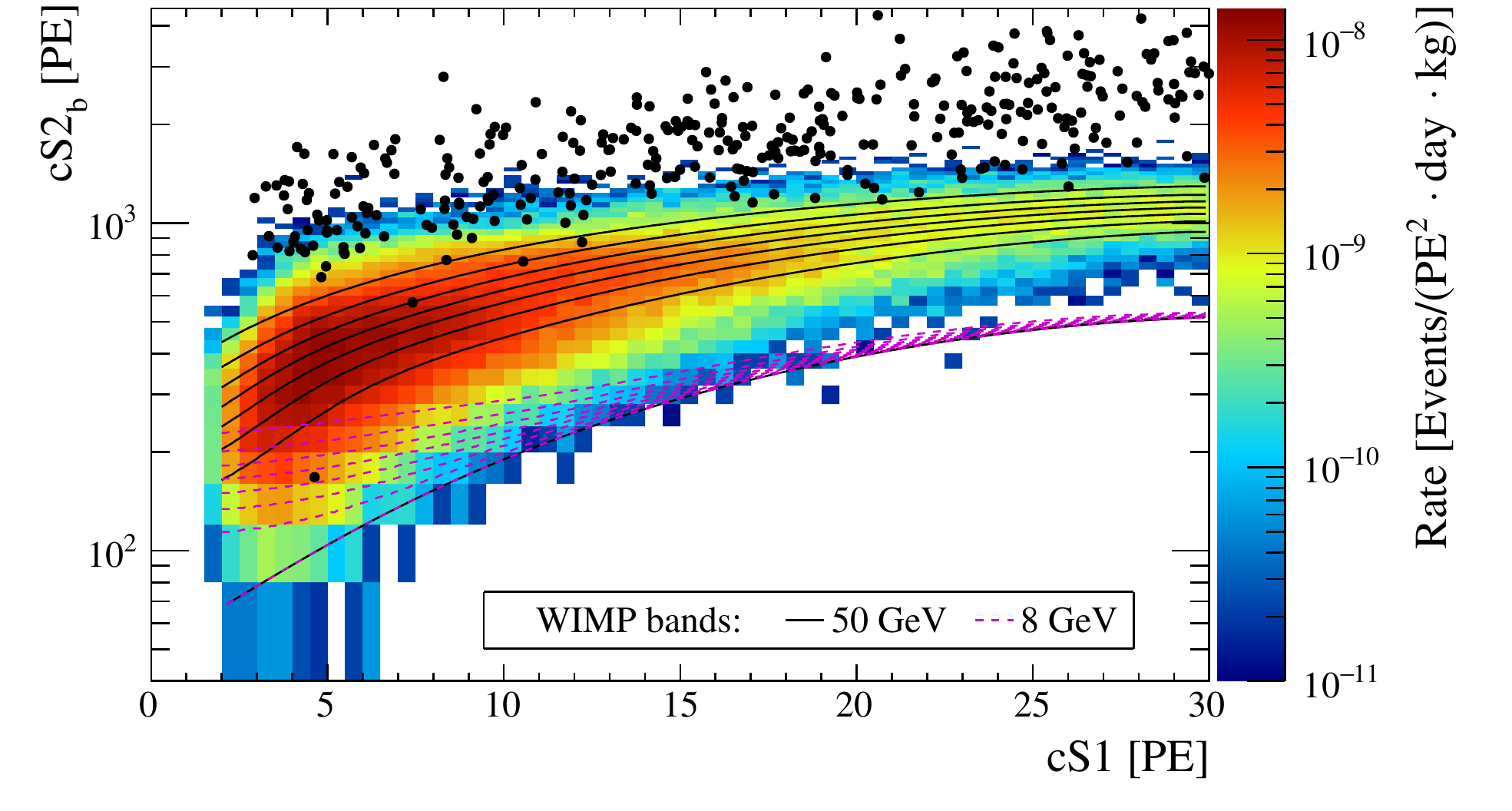}\vspace{-.1in}
   \caption[]{
   The ER Gaussian (top), non-Gaussian ER (middle), and NR (bottom) background predictions for run~III. The science data and signal bands, as in Fig.~\ref{Fig:pl_banding}, are overlaid for reference.
     \label{Fig:background_2d}}
  \end{center}	
\end{figure}

The NR background model is estimated by Monte Carlo 
simulation~\cite{Aprile:2013tov}, including a radiogenic 
component, $f^{NR}_{RG}$, from ambient materials and a cosmogenic component,
$f^{NR}_{CG}$, from cosmic radiation and their secondary processes. 
The computed energy spectra are translated to cS1 and cS2$_{\rm b}$ following the procedure in the previous Sec.~\ref{sec:signal} and normalized to the exposure of each run. 
The total NR background prediction is then $f^{NR} = f^{NR}_{RG} + f^{NR}_{CG}$,
where the functional dependence on cS1 and cS2 is suppressed for brevity, 
and shown in Fig.~\ref{Fig:background_2d} (bottom). 

The ER background consists of a Gaussian-shaped
component and a non-Gaussian component. 
The Gaussian component, $f^{ER}_{G}$ shown in Fig.~\ref{Fig:background_2d} (top), 
is modeled as in~\cite{Aprile:2012nq} by parametrizing
the ER calibration data from each run and normalizing to the dark matter
data above the ROI. 

\begin{figure}[h]
  \begin{center}
   \includegraphics[scale = 0.4]{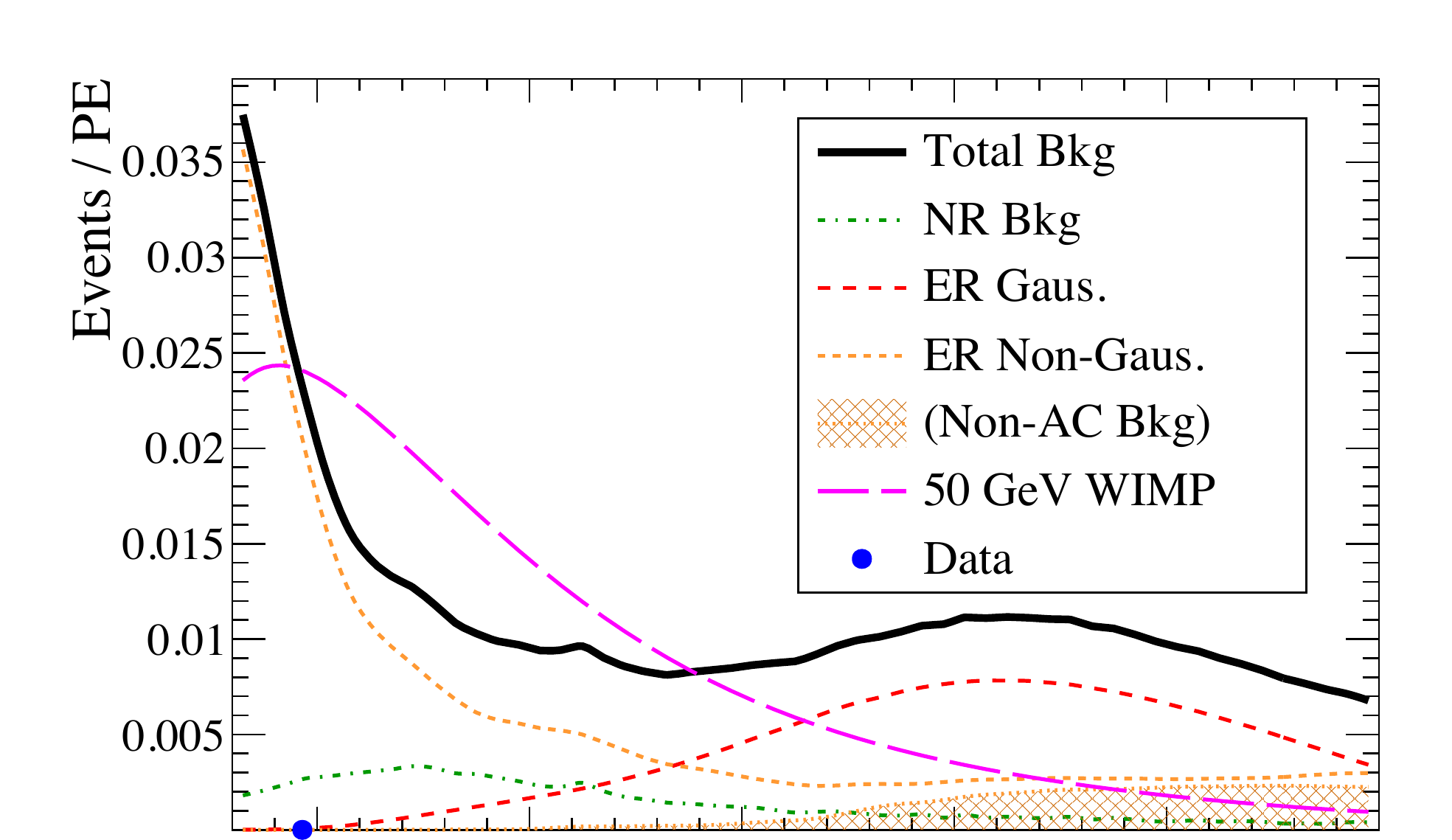}\vspace{-.0in}
   \includegraphics[scale = 0.4]{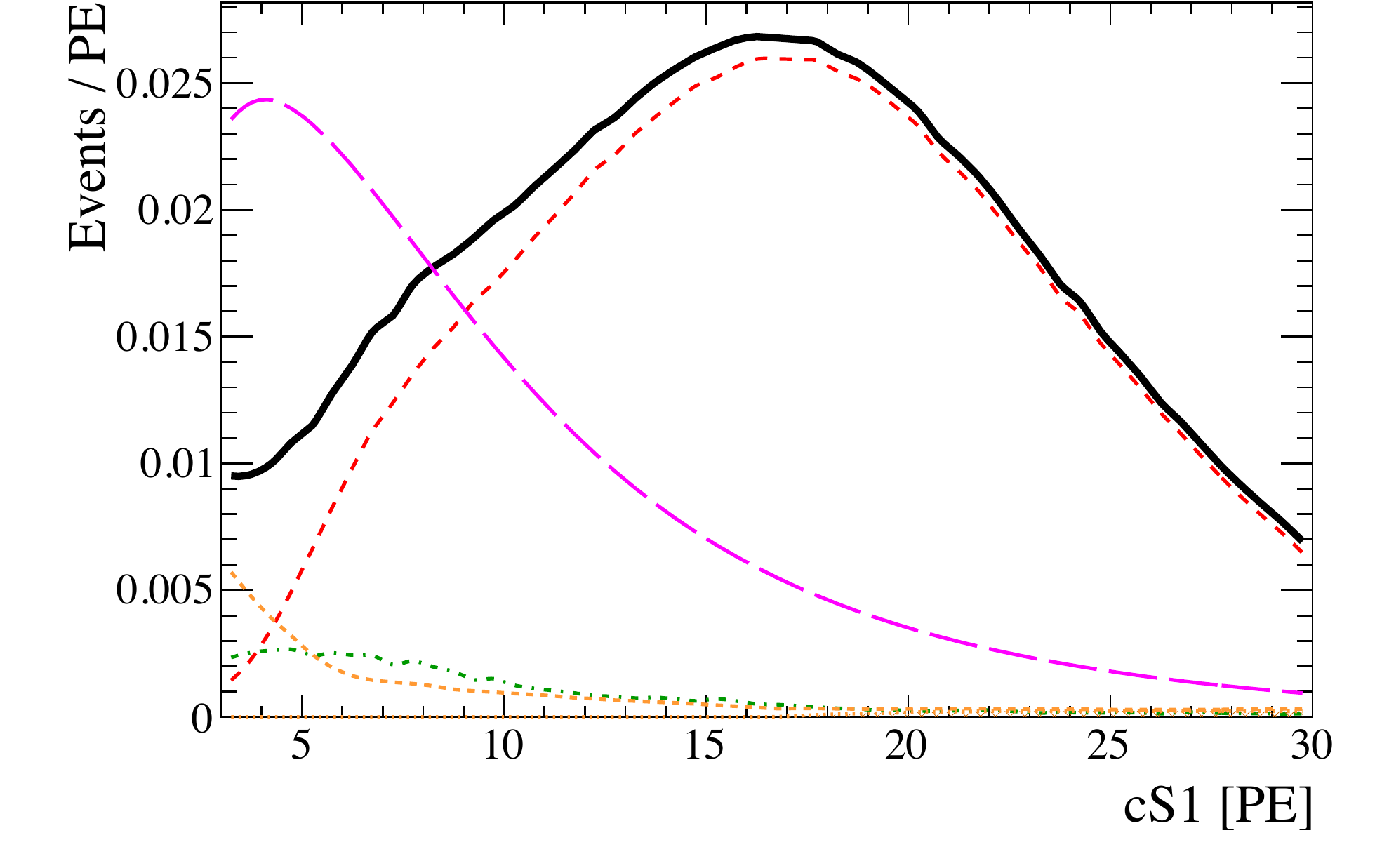}\vspace{-.1in}
   
   \caption[]{Expected event rates for PL band 0 (top) and 2 (bottom) from 
Fig.~\ref{Fig:pl_banding} for a 50~GeV/$c^2$ WIMP and an assumed SI cross section of $\sigma_{SI} = 10^{-45}~\mbox{cm}^{2}$ (long-dashed magenta line). The contribution from various 
background components described in the text (nonsolid colored lines) are shown together
with their sum (solid black line). The non-AC component of the total non-Gaussian ER background is shown (diagonal line filled area).
The run~III models are shown for example and the cS1 of the ROI event is shown on the 
horizontal axis (blue point).
   \label{Fig:models_banded}}
  \end{center}	
\end{figure}

The non-Gaussian component consists of anomalous events, 
such as those that show incomplete charge collection and
accidental coincidences (AC) of lone (uncorrelated) S1s and S2s. 
Previously~\cite{Aprile:2012nq}, 
these events were effectively modeled by a parametrization, $f^{ER}_{AN}$, of ER calibration events 
in the ROI after subtraction of the Gaussian component. However, this model
is underestimating the effect of the AC contribution. 
Hence, a more physically motivated procedure considering both non-Gaussian contributions is used to derive the background model. The new AC component model, described in the Appendix, 
identifies high statistics samples of lone S1s and S2s
to estimate this background with a better understanding of the spectral shape.
The product of the rates of these two samples gives the AC rate.
Distinct AC rates for both the ER calibration data, $f^{ER}_{AC}$, 
and dark matter data, $f^{DM}_{AC}$, can be derived using this method.
The prediction for ER calibration data is 
consistent with the observed number of events in the ROI, validating the model.
The total non-Gaussian model is then
given by $f^{ER}_{NG} = f^{DM}_{AC} + \max\left(f^{ER}_{AN} - f^{ER}_{AC}, 0\right)$, 
where the last term describes any remaining part of the anomalous leakage that is not accounted for by 
accidental coincidences. This model is shown in Fig.~\ref{Fig:background_2d} (middle), 
where the bulk at low S1 is dominated by the AC component, whereas the tail towards
high S1 can be explained by the non-AC anomalous leakage component.
The contribution of each component is shown for two example PL bands in Fig.~\ref{Fig:models_banded}.

\begin{table}
\caption{Relative contribution (\%) of each background component in the ROI. \label{table2}}
\centering
\begin{tabular}{|l|c|c|c|} \cline{1-4}
\multicolumn{1}{|l|}{}    & Run~I & Run~II & Run~III  \\ \cline{1-4}
Gaussian ER &  $64\pm 6$ & $55\pm 8$  & $72\pm 7$   \\ 
Non-Gaussian ER & $33\pm 5$ & $35\pm 7$ & $19\pm 4$  \\ 
NR & $3\pm 2$ & $10\pm 7$ & $9\pm 7$ \\ \cline{1-4}
\end{tabular}
\end{table}

\begin{figure}[h]
  \begin{center}
   \includegraphics[scale = 0.4]{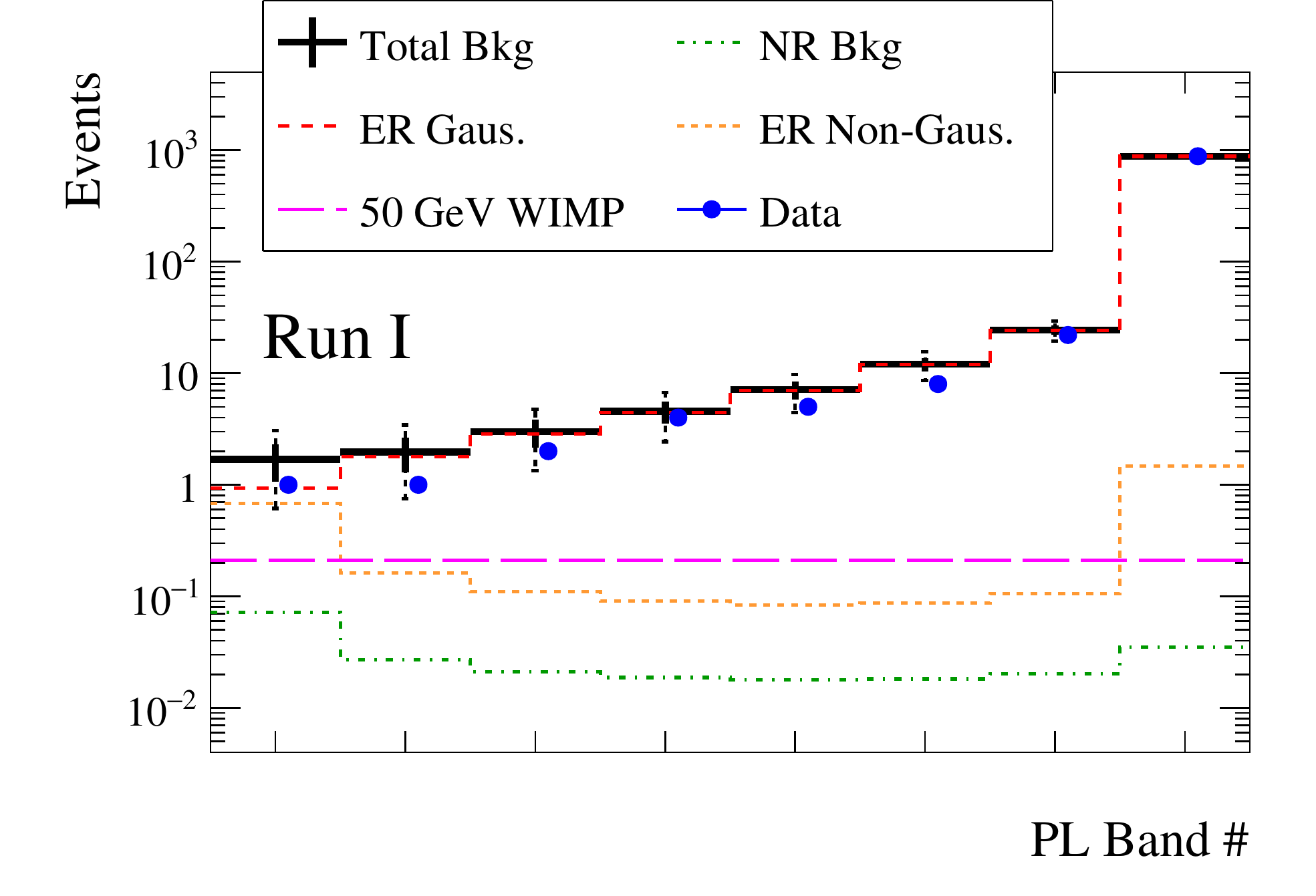}\vspace{-.34in}
   \includegraphics[scale = 0.4]{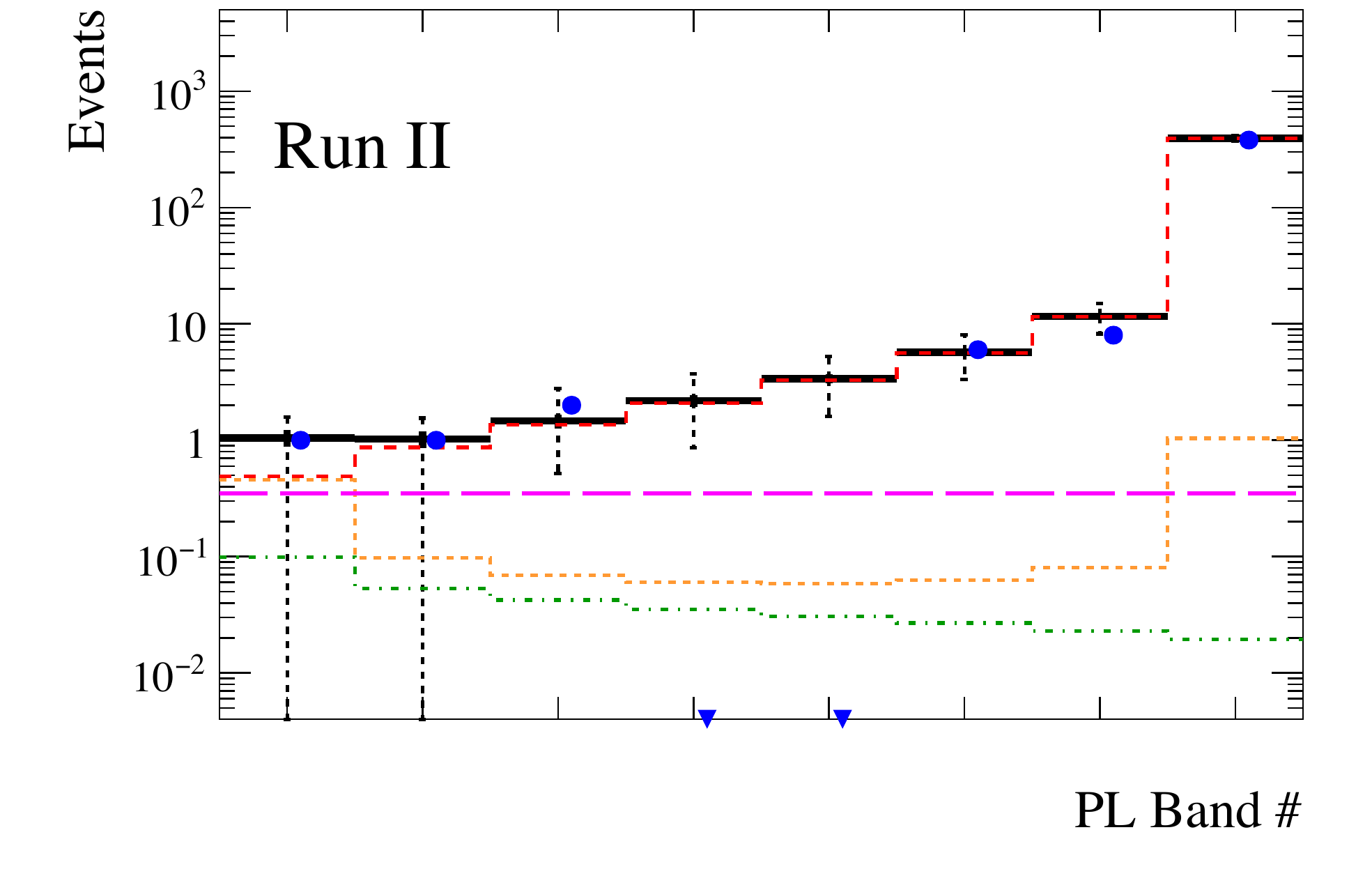}\vspace{-.34in}
   \includegraphics[scale = 0.4]{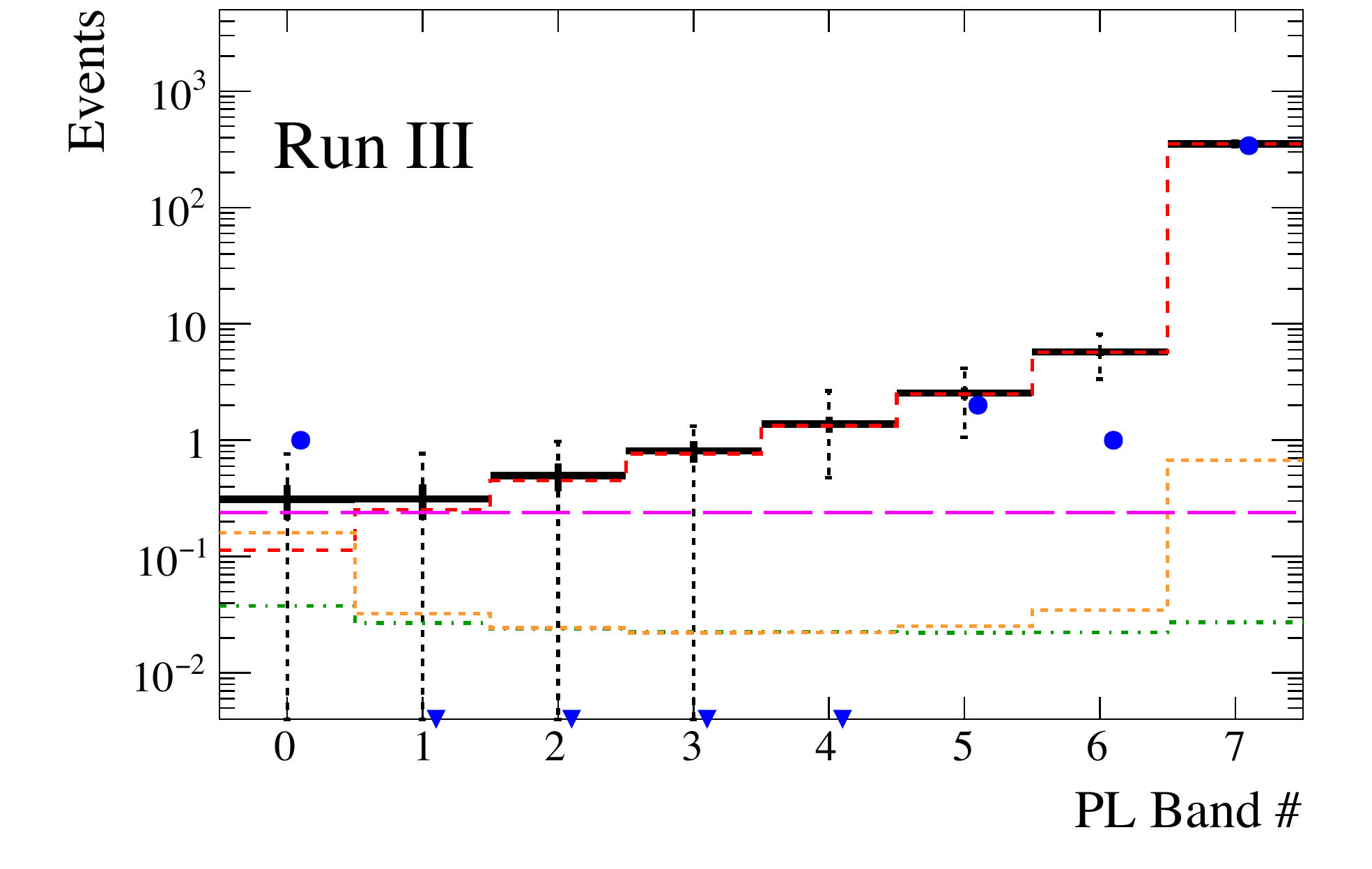}\vspace{-.1in}
  \end{center}	
   \caption[]{Integrated event rates for each PL band assuming a 50~GeV/$c^2$ WIMP 
   at $\sigma_{SI} = 10^{-45}$~cm$^2$ in runs I (top), II (middle), and 
   III (bottom). Banding and legend similar to 
   Fig.~\ref{Fig:models_banded}. The solid line error bars on the total correspond to the total Poisson error defined from ER calibration data shown in Fig.~\ref{Fig:models_errors}, while the dotted lines show the 68\% Poisson probability region for the expectation. \label{Fig:models_integrated}}
\end{figure}

Finally, the total background model is given by
\begin{equation}
f_{b} = f^{NR} + f^{ER}_{G} + f^{ER}_{NG},
\label{Eq:Nb}
\end{equation}
for each run, shown in Fig.~\ref{Fig:pl_banding} for run~III. The projection 
in cS1 for two example bands is shown in Fig.~\ref{Fig:models_banded} 
including the contribution from each background component.
The integrated event rate for each PL band is shown in 
Fig.~\ref{Fig:models_integrated} and the fractional contributions to the ROI for 
each run are shown in Table~\ref{table2}. Run~I is $^{85}$Kr dominated which results in a smaller relative contribution of the NR background in comparison to runs~II and III. 
The non-Gaussian data-driven model predicts a smaller contribution in run~III compared to run~II.
A sideband unblinding of the run~III science data around the ROI
was performed similarly to run~II~\cite{Aprile:2012vw} to test and validate
the background models. No significant deviations from the predictions were found.

\begin{figure}[h]
  \begin{center}
   \includegraphics[scale = 0.4]{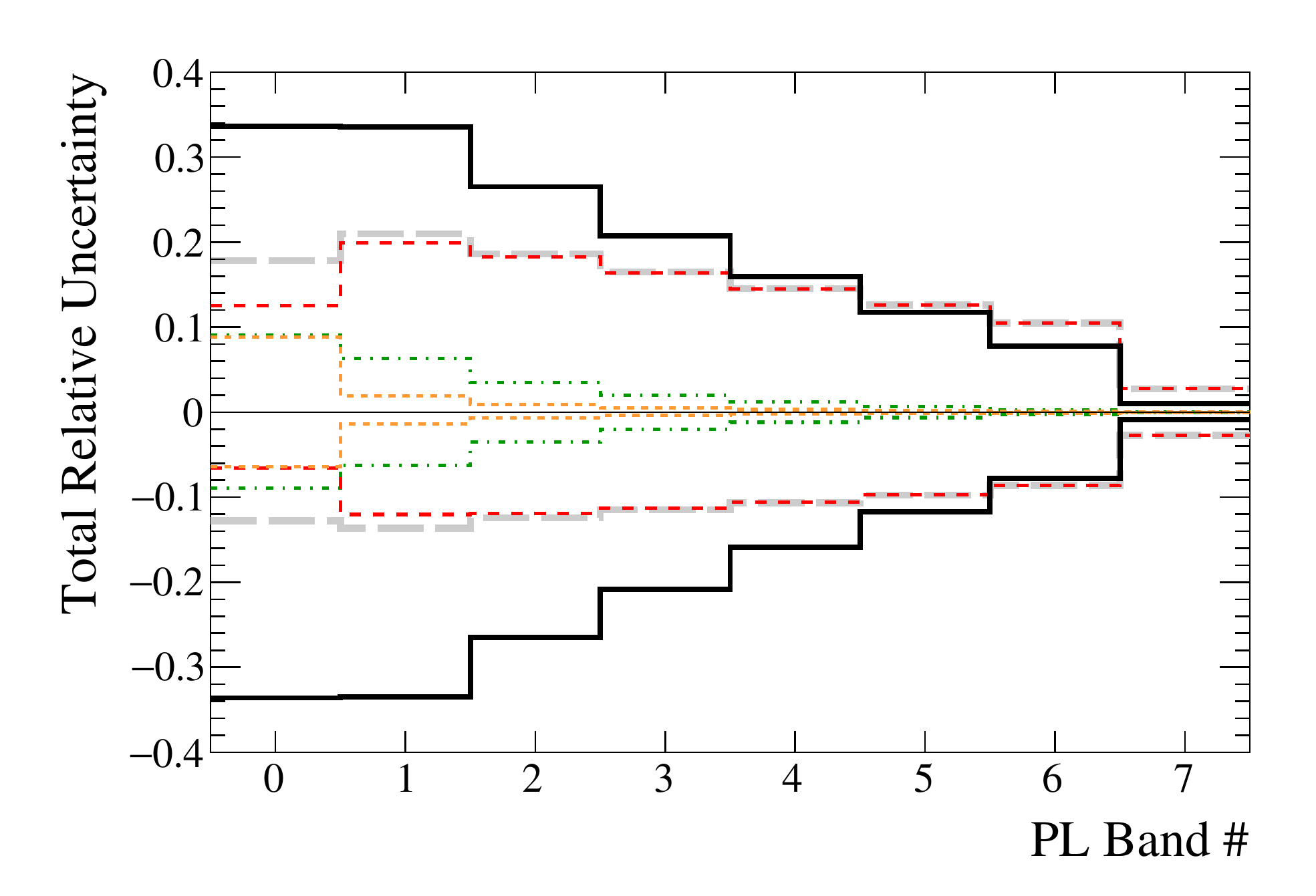}
   \caption[]{Total uncertainty for each background component and their 
   quadrature sum (long-dashed gray line) in run III. Banding and legend similar to Fig.~\ref{Fig:models_banded}, except for
   including the Poisson error defined from ER calibration data (solid black line) to visualize the 
   constraint term in Eq.~(\ref{eq:ercal_pois}). Positive and negative lines show an asymmetric uncertainty.
   \label{Fig:models_errors}}
  \end{center}	
\end{figure}

The previous PL analysis~\cite{Aprile:2011hx} assumed an effective uncertainty on the total background model
by including a Poisson constraint term based on the number of ER calibration events 
in each band (Eq.~(\ref{eq:ercal_pois}) below). This uncertainty is now cross-checked by propagating 
the systematic errors for each background component, including errors from the 
parametrization fits to calibration data, selection criteria and efficiency uncertainties 
for the AC model, and muon flux normalization uncertainty for the NR component.
The total error for each background component and their quadrature sum 
is shown in Fig.~\ref{Fig:models_errors}. The Poisson error is chosen for this analysis 
as it conservatively overestimates the propagated errors, which may be
overconstrained from the assumed (nonphysical) parametrizations.

\subsection{\label{sec:likelihood} The likelihood function}

The signal hypothesis test is performed by means of a profiled likelihood ratio test statistic 
and its asymptotic distributions~\cite{Cowan:2010js}.
The procedure is described in detail in~\cite{Aprile:2011hx} 
and only the modifications for this analysis are highlighted here. 


The full likelihood for the combination of the three science runs 
can be written as:
\begin{equation}
\mathcal{L} =  \mathcal{L}^{I} \times \mathcal{L}^{II} \times \mathcal{L}^{III} \times \mathcal{L}_3(t_{\mathcal{L}_{\mathrm{eff}}}) \times \mathcal{L}_4 (t_{{Q}_{y}}), 
\end{equation}
where the likelihood function for a given science run, $i$, is
\begin{equation}
\mathcal{L}^{i} = \mathcal{L}_1^i (m_{\chi}; \sigma, N^{i} _{b}, \bm{\epsilon ^{i} _{b}}
, t_{\mathcal{L}_{\mathrm{eff}}}, t_{{Q}_{y}})  \times \mathcal{L}_2 ^i(\bm{\epsilon _{b} ^i}).
\end{equation}
where $\bm{\epsilon _{b}^i}$ indicates a vector of the background nuisance parameter per band $j$ and 
\begin{equation}
\begin{split}
\mathcal{L}_1^i = \prod ^{K^i (m_{\chi})} _j \mathrm{Poiss}\left( n^{i,j}|\epsilon^{i,j} _{s}N^{i}_s(\sigma)+\epsilon^{i,j} _{b} N^{i} _{b} \right) \times \\
\prod _{k=1} ^{n^{i,j,k}}\frac{ \epsilon^{i,j}_{s} N^{i}_s(\sigma) f_s^{i,j}(\mathrm{cS1}^k) + \epsilon^{i,j}_{b} N^{i} _{b} f_{b}^{i,j}(\mathrm{cS1}^k)}{ \epsilon^{i,j}_{s}N^{i}_s(\sigma) + \epsilon^{i,j}_{b} N^{i}_{b}}
\label{Eq:L1}
\end{split}
\end{equation}
is the extended likelihood function. The number of observed events is $n^{i,j}$,
and $N^{i}_s$ and $N^{i}_\mathrm{b}$ are 
the maximum likelihood estimators (MLEs) 
 for the total number of signal and background events, 
respectively. 
The ROI is divided into 8 bands, 
$K^i (m_{\chi})$, depending on the WIMP mass as depicted in Fig.~\ref{Fig:pl_banding}. 
The fractions, $\epsilon^{i,j}_{s,b}$, for each band are derived from
the signal and background models. 
$N^{i}_s(\sigma)$ is related to the cross section of interest, $\sigma$, 
via Eq.~(\ref{Eq:Ns}).
The dependencies of $N^{i}_s$, $\epsilon^{i,j}_{s}$, and $f_s^i$ on 
$t_{\mathcal{L}_{\mathrm{eff}}}$, $t_{{Q}_{y}}$, and LCE are suppressed for clarity.
 The shapes in cS1, $f_{s,b}$, are considered for each event, $k$,
in the second term of Eq.~(\ref{Eq:L1}). 
The background model uncertainties, shown in Fig.~\ref{Fig:models_errors} (black line), 
are modeled through variations of $\epsilon^{i,j}_{b}$, constrained by
\begin{equation}
\mathcal{L}_2^i = \prod ^{K^i(m_{\chi})} _j \mathrm{Poiss}(m^{i,j}_{b}|\epsilon^{i,j} _{b} M^i_{b}),
\label{eq:ercal_pois}
\end{equation}
where $M^i_{b}$ is the total number of ER calibration events and $m^{i,j}_{b}$ 
is the number in each band. The global nuisance parameters 
$t_{\mathcal{L}_{\mathrm{eff}}}$ and $t_{{Q}_{y}}$
are constrained by external light and charge yield measurements through
\begin{equation}
\mathcal{L}_{3,4}(t_{\mathcal{L}_{\mathrm{eff}}}, t_{Q_y}) = exp(-(t_{\mathcal{L}_{\mathrm{eff}}},t_{Q_y})^2/2),
\end{equation}
with the allowed variation derived from the spread and uncertainties 
in those data~\cite{Aprile:2011hi}.
 
\section{WIMP Search Results}
\label{sec:results}

\begin{figure}[h]
  \begin{center}
   \includegraphics[scale = 0.47]{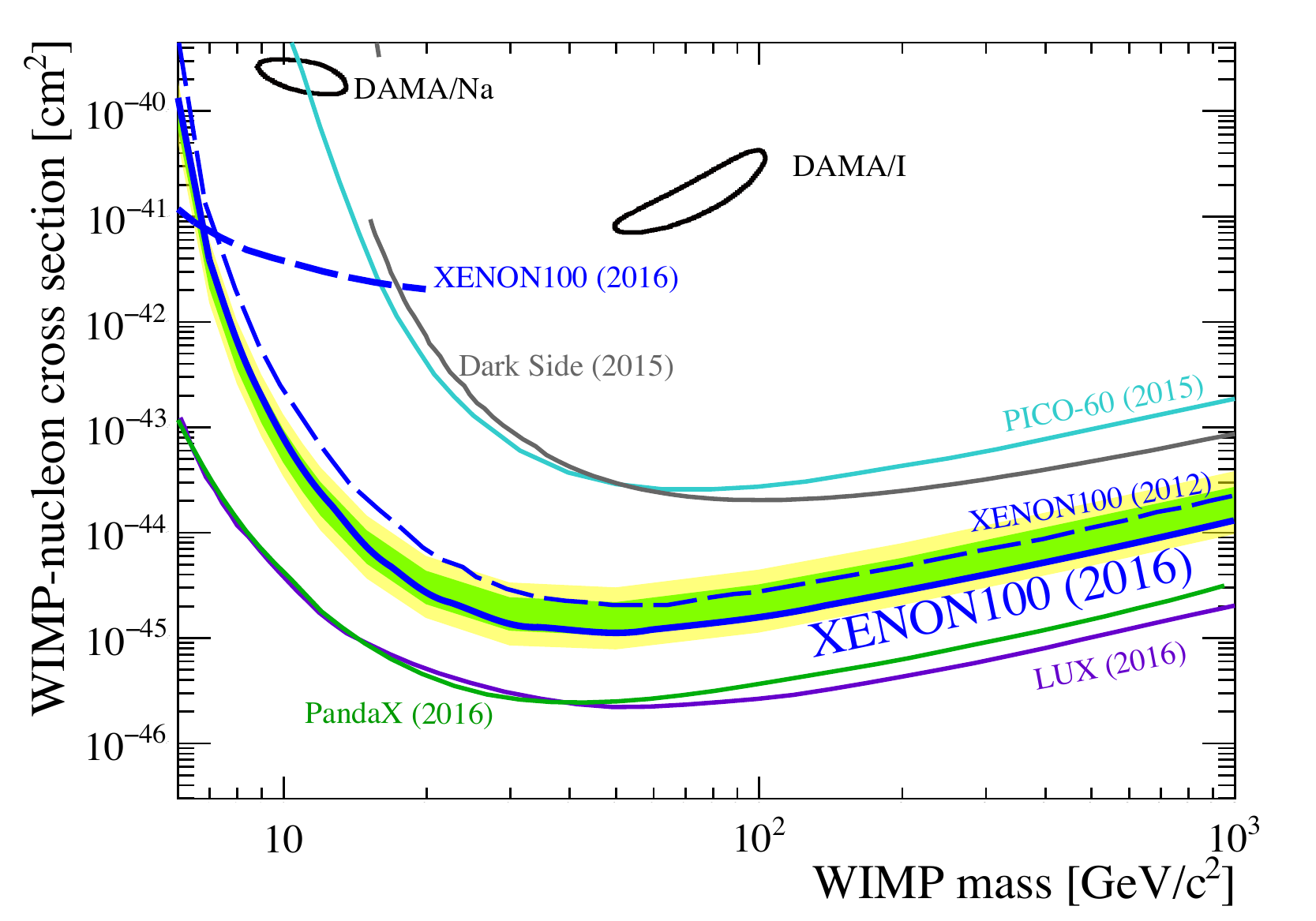}
   \caption[]{Spin-independent cross section limit (blue line) and 
   1$\sigma$ (green band) and 2$\sigma$ (yellow band) expected sensitivity 
   regions at 90\% C.L. from the combined analysis of the three XENON100 
   science runs. For comparison, a subset of other experimental limits (90~\%C.L.) 
   and detection claims (2$\sigma$) are also shown \cite{Bernabei:2008yi,Aprile:2012nq,Amole:2015pla,Aprile:2016wwo,Tan:2016zwf,Akerib:2016vxi,Agnes:2015ftt}.
    \label{Fig:results_si}}
    

  \end{center}	
\end{figure}

After unblinding the run III ROI, no significant excess of events over 
the expected background is observed, as shown in Table~\ref{table1} and Fig.~\ref{Fig:models_integrated}.
The PL analysis of the combined data results in a 90\% confidence level (C.L.) limit using the C.L.$_\mathrm{s}$ 
prescription~\cite{Read:2000} on the WIMP-nucleon SI cross section
as shown in Fig.~\ref{Fig:results_si}, corresponding 
to $1.1\times10^{-45}$~cm$^2$ at a 50~GeV/$c^2$ mass.
The green and yellow sensitivity bands 
represent the distribution of expected upper limits under the assumption of no signal. 
A cross-check with a second independent PL code using the same inputs,
as well as an order of magnitude check with a maximum gap analysis~\cite{Yellin:2011xf}, 
resulted in limits consistent within the sensitivity bands.
The XENON100 run~III result confirms the absence of 
a WIMP dark matter signal and a combination of the data improves the limit 
on the SI WIMP-nucleon cross section by a factor of 1.8 at 50~GeV/$c^2$ mass
compared to the previously published XENON100 limit~\cite{Aprile:2012nq}.

\begin{figure}[h]
  \begin{center}
   \includegraphics[scale = 0.45]{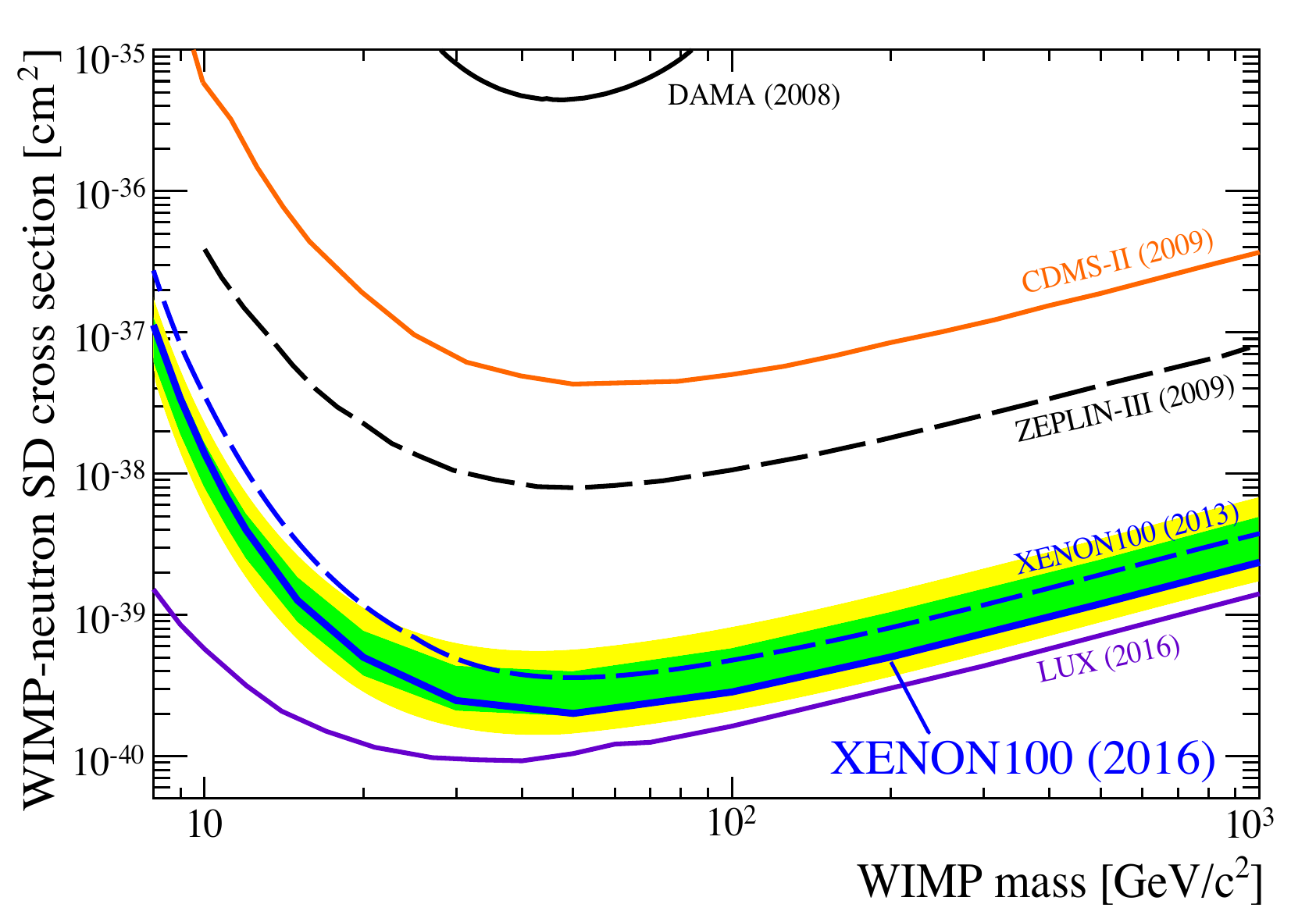}\vspace{-0.25cm}
    \includegraphics[scale = 0.45]{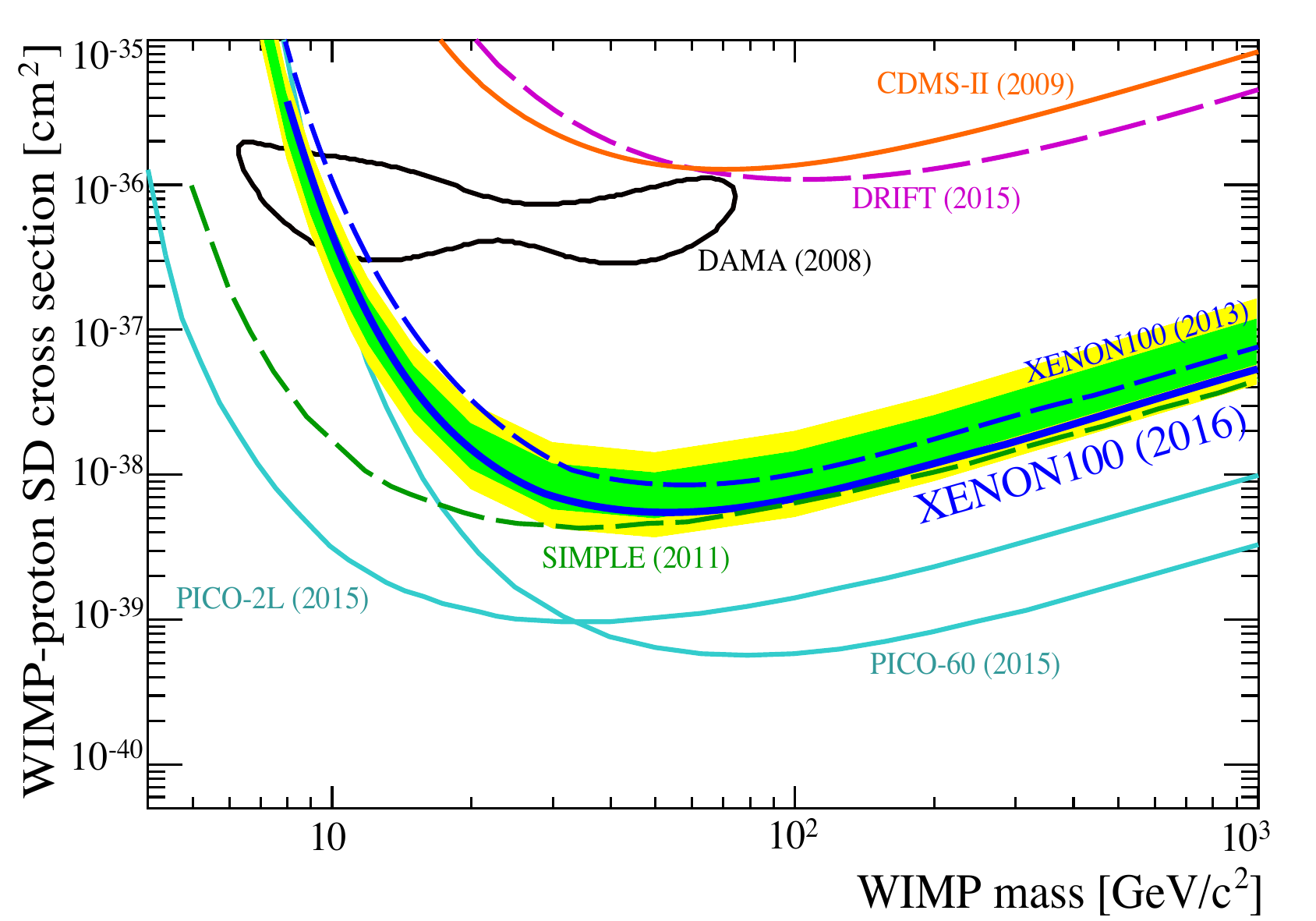}
   \caption[]{Spin-dependent cross section limit (blue line) and 
   1$\sigma$ (green band) and 2$\sigma$ (yellow band) expected sensitivity 
   regions at 90\% C.L. from the combined analysis of the three XENON100 
   science runs. The top (bottom) 
   panel shows the individual neutron (proton) only cross sections. For comparison, other experimental limits (90~\%C.L.)
   and detection claims (2$\sigma$) are also shown~\cite{Akerib:2016lao,Amole:2015pla,Amole:2016pye,Battat:2014van,Aprile:2013doa,Bernabei:2001ve,Lebedenko:2009xe,Ahmed:2008eu,Felizardo:2011uw}.
   \label{Fig:results_sd}}
   


  \end{center}	
\end{figure}

We apply the same statistical approach to set upper limits on the 
SD WIMP-proton and neutron cross sections, shown in Fig.~\ref{Fig:results_sd}. 
For coupling to protons, the limit at 50~GeV/$c^2$ is 5.2$\times 10^{-39}$~cm$^2$, 
whereas for neutrons it is 2.0$\times 10^{-40}$~cm$^2$. This constitutes improvements by factors of 
1.7 and 1.8, respectively, compared to the previously published XENON100 limits~\cite{Aprile:2013doa}.

\section{Summary}
We present the final XENON100 spin-independent and spin-dependent results from the
combined analysis of two already published science runs and a third new run, with
a total exposure of 477 live days (48~kg$\times$yr) acquired between January 2010 and 
January 2014. Improvements to the data quality event selection were described, resulting 
in a reduction of background and increase in purity of the final dark matter sample. 
A new technique to quantify accidental coincidences was developed and implemented into
the ER background model. Furthermore, the signal model is now computed analytically 
for S1 and S2, including more accurate modeling of all acceptances and thresholds. 
Finally, requiring a minimum number of detected signal quanta improves the robustness 
of the analysis close to the energy threshold, which is important for low WIMP masses.
No evidence for dark matter is found and an upper limit of the WIMP-nucleon cross 
section is derived. The combination of the three science runs with the improved analysis 
results in a SI limit of $1.1\times10^{-45}$~cm$^2$ at a 50~GeV/$c^2$ mass and a SD 
neutron (proton) limit of $2.0\times10^{-40}$~cm$^2$ ($5.2\times10^{-39}$~cm$^2$) 
at 50~GeV/$c^2$ mass.




\begin{acknowledgments}
We gratefully acknowledge support from the National
Science Foundation; Swiss National Science Foundation;
Deutsche Forschungsgemeinschaft; Max Planck
Gesellschaft; Foundation for Fundamental Research on
Matter; Weizmann Institute of Science; Israeli Centers Of Research Excellence; Initial
Training Network Invisibles (Marie Curie Actions, PITNGA-2011-289442);
Fundacao para a Ciencia e a Tecnologia;
Region des Pays de la Loire; Knut and Alice Wallenberg
Foundation; and Istituto Nazionale di Fisica Nucleare.
We are grateful to Laboratori Nazionali del Gran
Sasso for hosting and supporting the XENON project.
\end{acknowledgments}

\begin{appendix}
\section{Accidental Coincidence Background Model}
\label{app:ac}

\begin{figure}[h]
  \begin{center}
   \includegraphics[scale = 0.45]{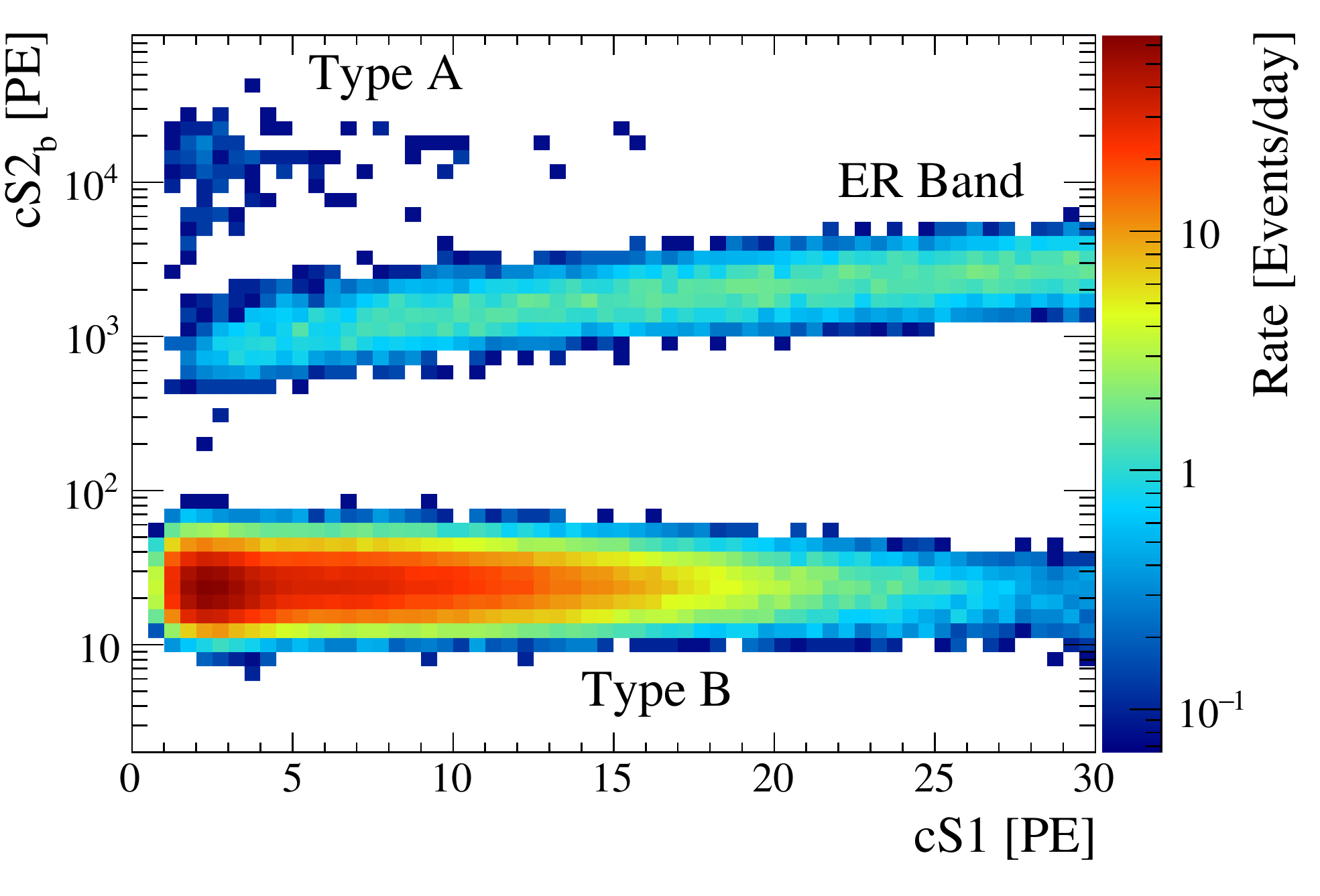} \vspace{-.5cm}
   \caption[]{Event categorization (types A and B) for the accidental
   coincidence background model. The run~III ER calibration data
   is shown, but is similar for all runs and DM science data.
   \label{Fig:ac_lone_samples}}
  \end{center}	
\end{figure}
\end{appendix}

A data-driven method to estimate the accidental coincidence (AC) rate was developed, similar to~\cite{PhysRevD.92.052004}.
Lone S2s are selected with the same S2-related criteria, 
referenced and described in Sec.~\ref{sec:cuts}, as well as 
requiring no S1 preceding the S2 in the event waveform. 
To derive the lone S1 spectrum, events in the S2-S1 plane 
are categorized into two regions that are known to consist mostly 
of ACs: type~A events with a large S2 paired 
with a small S1, and type~B events with a very small S2 paired with 
any S1, as shown in Fig.~\ref{Fig:ac_lone_samples}.
Type~A events are mostly ACs, but are limited to lower S1s and in statistics. Type~B events have large statistics across S1, 
but are contaminated by events where the S2 was caused by the S1
through impurity photoionization S2s.
These secondary S2s are modeled by the rate difference between type~A 
and type~B events. The lone S1 spectrum is then derived from the
type~B S1 spectrum after subtracting the secondary S2s. Finally, 
the AC rate is given by the product of the lone S1 and lone S2 spectra, 
and is shown in Fig.~\ref{Fig:background_2d} (middle).
The uncertainty, shown in Fig.~\ref{Fig:models_errors}, is dominated by 
systematic uncertainties from the modeling of the secondary S2s which 
is limited by type~A event statistics.

\bibliography{main}

\end{document}